\documentclass{aa}
\usepackage{amsmath}
\usepackage{nicefrac}
\usepackage{siunitx}
\usepackage{listings}
\usepackage{color}
\usepackage{bm}
\usepackage{graphicx}
\usepackage[utf8]{inputenc} 
\def\aap{A\&A}

\def\apjs{ApJS} 
\def\apjl{ApJ}

\newcommand{\vect}[1]{\bm{\mathrm{#1}}}

\usepackage{multirow}
\newlength{\digit}
\newcommand{\tc}[1]{\multicolumn{1}{c}{{#1}}}
\newcommand\tstrut{\rule{0pt}{2.5ex}}

\usepackage{tipa}
\begin{document}

\title{A smooth particle hydrodynamics code to model collisions between solid, self-gravitating objects}

\author{C.~Schäfer\thanks{
If you are interested in our CUDA SPH code {\tt miluphCUDA}, please write an email to Christoph Schäfer. 
{\tt miluphCUDA} is the CUDA port of {\tt miluph}.
{\tt miluph} is pronounced [\textipa{ma\i l\textturnv v}].
\newline
We do not support the use of the code for military purposes.
}\inst{1} \and S.~Riecker\inst{1} \and T.I.~Maindl\inst{2} \and R.~Speith\inst{3}
     \and S.~Scherrer\inst{1} \and W.~Kley\inst{1}
}

\institute{Institut für Astronomie und Astrophysik, Eberhard Karls Universität Tübingen, Auf der Morgenstelle 10, 72076
Tübingen \email{ch.schaefer@uni-tuebingen.de}
\and
Department of Astrophysics, University of Vienna, Türkenschanzstraße 17, 1180 Vienna 
\and
Physikalisches Institut, Eberhard Karls Universität Tübingen, Auf der Morgenstelle 14, 72076 Tübingen}

\date{Received ---- ; accepted ----}

\abstract
{Modern graphics processing units (GPUs) lead to a major increase in the performance of the computation of astrophysical
simulations. Owing to the different nature of GPU architecture compared to traditional central processing units (CPUs) such as x86
architecture, existing numerical codes cannot be easily migrated to run on GPU.
Here, we present a new implementation of the numerical method smooth particle hydrodynamics (SPH) using 
CUDA\texttrademark\ and the first astrophysical application of the new code: the collision between Ceres-sized objects.} 
{The new code allows for a tremendous increase in speed of astrophysical simulations with SPH and self-gravity at low costs for
new hardware.}
{We have implemented the SPH equations to model gas, liquids and elastic, and plastic solid bodies and added a
fragmentation model for brittle materials. Self-gravity may be
optionally included in the simulations and is treated by the use of a Barnes-Hut tree.}
{We find an impressive performance gain using NVIDIA consumer devices compared to our existing OpenMP code. The new code
is freely available to the community upon request.}
{}
\titlerunning{A SPH Code to Model Collisions}
\authorrunning{Schäfer et al.}
\maketitle


\section{Introduction} 
Since the introduction of the first CUDA (formerly known as Compute Unified Device Architecture) development kit by the
NVIDIA Corporation and even more since the support of double precision by modern GPU hardware (i.e., NVIDIA GT200 chip
with compute capability of $1.3$ and higher), many astrophysical applications that make use of the acceleration by
general-purpose computing on graphics processing units have been developed
(e.g.~\citealt{2012JCoPh.231.2825B,2014ApJS..214...12K}). The modern GPUs allow for a significant performance gain
compared to serial codes for central processing units (CPUs). The lower cost of GPUs compared to high performance
computing clusters of CPUs leads to higher performance per cost in favor of the GPU. Moreover, GPUs are
inherently more energy efficient than CPUs because they are optimized for throughput and performance per watt rather than for 
absolute performance \citep{tingxing:2015}. Henceforth, astrophysical problems that in general demand cluster
hardware may even be addressed on workstations with NVIDIA hardware. 

We decided to implement a CUDA port of our parallelized OpenMP SPH code to benefit from these modern GPUs. In this
publication, we present the physical models, their CUDA implementation, the validation simulations of our new code, and
we show a first glimpse at possible applications. The SPH code may be used for hydrodynamical simulations and to model
solid bodies, including ductile and brittle materials. The code uses a tree code algorithm for the additional treatment
of self-gravity. We show several validation simulations: the simulation of elastic rubber cylinders, the gravitational
collapse of a molecular cloud, the impact of an aluminium bullet into an aluminium cube and the impact of a nylon bullet
in a basaltic rock.

Extended hydrocodes have been applied to model large deformation and strong shock wave physics using different
numerical schemes. The Eulerian grid code CTH has several models that are useful for simulating strong shock, large
deformation events, and this code can be used to model elastic-plastic behavior and fracture \citep{MCGLAUN1990351}. The iSALE grid
code is a comprehensive tool to study impacts \citep{2006Icar..180..514W, 2009Icar..204..716E}. Apart from grid codes,
there are also particle-based numerical schemes.
Several SPH codes for modern cluster hardware are available to the community. The well-known Gadget-2 code by
\cite{2005MNRAS.364.1105S} may be the most sophisticated and most widely applied SPH code for the simulation of
astrophysical flows. The SPH code by
\cite{1994Icar..107...98B} with fundamental improvements of the physical models later by \cite{2008Icar..198..242J} has been
successfully applied to simulate impacts and collisions of astrophysical objects. However, there is no freely
available SPH code for the simulation of solid and brittle material including the treatment of self-gravity at hand.
Our new code bridges this gap and high resolution SPH simulations become feasible and may be performed even on a
standard workstation with a GPU from the consumer price range. 

The outline is as follows. In the next section we present the physical models and the governing equations that are
solved by the code. In sect.~\ref{section:numerical_model} we describe the smooth particle hydrodynamics numerical model
for the simulation of solid bodies, the Grady-Kipp damage model for the formation of cracks and stress weakening, and
the Barnes-Hut tree algorithm for gravitational forces.  We give a detailed description of our implementation of these
algorithms in the new CUDA code.  We present the results of three standards tests for the code in
sect.~\ref{section:tests}: the collision of two perfect elastic rubber cylinders, the high-velocity impact of a bullet
into an aluminium cube, and the collapse of an isothermal molecular cloud.  Our first application of the new code s
described in sect.~\ref{section:application} and we present its results in sect.~\ref{section:results} where we
additionally discuss the limitations of the current code version in regard to numerical issues as well as in terms of
the physical models. Eventually, we conclude in sect.~\ref{section:conclusion}.

\section{Physical model\label{section:physical_model}}
In this section, we present the set of partial differential equations from continuum mechanics in Cartesian coordinates
that we apply to model gas and solid bodies.
Throughout this paper, the Einstein summation rule is used, Greek indices denote spatial coordinates and run from 1 to
3.

\subsection{Mass conservation}
The equation of continuity is the equation of the conservation of mass. It is given in Lagrangian form by
\begin{equation}
\label{eq:conservation_of_mass}
\frac{\mathrm{d} \varrho}{\mathrm{d}t} + \varrho \frac{\partial v^\alpha}{\partial x^\alpha} = 0,
\end{equation}
where $\varrho$ denotes the density and $\vect{v}$ the velocity of the solid body or fluid.
\subsection{Conservation of momentum}
The equation of the conservation of momentum in continuum mechanics for a solid body reads in Lagrangian formulation
\begin{equation}
\label{eq:conservation_of_momentum}
\frac{\mathrm{d} {v^\alpha}}{\mathrm{d}t} =  \frac{1}{\varrho}  \frac{\partial \sigma^{\alpha
\beta} }{\partial x_\beta},
\end{equation}
where $\sigma^{\alpha\beta}$ is the stress tensor given by the pressure $p$ and the deviatoric stress tensor
${S}^{\alpha \beta}$
\begin{equation}
\sigma^{\alpha \beta} = -p \delta^{\alpha \beta} + S^{\alpha \beta},
\end{equation}
with the Kronecker delta $\delta^{\alpha \beta}$.
For a liquid or gas, the deviatoric stress tensor vanishes and the conservation of momentum is given by the Euler
equation
\begin{equation}
\frac{\mathrm{d} {v^\alpha}}{\mathrm{d}t} =  - \frac{1}{\varrho}  \frac{\partial}{\partial x_\beta} p \delta^{\alpha \beta}.
\end{equation}
\subsection{Conservation of internal energy}
The equation of the conservation of the specific internal energy $u$ for an elastic body is given in Lagrangian formulation by
\begin{equation}
\label{eq:conservation_internal_energy}
\frac{\mathrm{d} u }{\mathrm{d} t} = - \frac{p}{\varrho} \frac{\partial v ^\alpha}{\partial x^\alpha} + \frac{1}{\varrho}
S^{\alpha \beta} \dot{\varepsilon}^{\alpha \beta},
\end{equation}
with the strain rate tensor $\dot{\varepsilon}^{\alpha \beta}$ given by eq.~(\ref{eq:strain_rate_tensor}).
\subsection{Constitutive equations}
In contrast to fluid dynamics, the set
(\ref{eq:conservation_of_mass};\ref{eq:conservation_of_momentum};\ref{eq:conservation_internal_energy}) of partial
differential equations is not sufficient to describe the dynamics of an elastic solid body, since the time evolution of
the deviatoric stress tensor $S^{\alpha \beta}$ is not yet specified. For the elastic behavior of a solid body, the
constitutive equation is based on Hooke's law. Extended to three dimensional deformations, it reads
\begin{equation}
S^{\alpha \beta} \sim 2 \mu \left( \varepsilon^{\alpha \beta} - \frac{1}{3} \delta^{\alpha \beta} \varepsilon^{\gamma
\gamma} \right).
\end{equation}
Here, $\mu$ denotes the material dependent shear modulus and $\epsilon^{\alpha \beta}$ is the strain tensor.
For the time evolution, it follows (for small deformations)
\begin{align}
\label{eq:stresstensor_evolution}
\frac{\mathrm{d}}{\mathrm{d}t}S^{\alpha \beta} =  2\mu \left( \dot{\varepsilon}^{\alpha \beta} - \frac{1}{3}
\delta^{\alpha \beta} \dot{\varepsilon}^{\gamma \gamma} \right)   &  \\  & +  \mathrm{rotation~terms,} \nonumber
\end{align}
where $\dot{\varepsilon}^{\alpha \beta}$ denotes the strain rate tensor, given by
\begin{equation}
\label{eq:strain_rate_tensor}
\dot{\varepsilon}^{\alpha \beta} = \frac{1}{2} \left( \frac{\partial v^\alpha}{\partial x^\beta} + \frac{\partial
v^\beta}{\partial x^\alpha} \right).
\end{equation}
The rotation terms in eq.~(\ref{eq:stresstensor_evolution}) are needed since the constitutive equations have to be
independent from the material frame of reference. There are various possibilities to achieve this. The usual approach
for particle codes is the Jaumann rate form (see, e.g.,\ \citealt{gray:2001} and references therein). The rotation terms
of the Jaumann rate form are
\begin{equation}
\label{eq:rotation_terms}
S^{\alpha \gamma} R^{\gamma \beta} - R^{\alpha \gamma} S^{\gamma \beta},
\end{equation}
with the rotation rate tensor
\begin{equation}
R^{\alpha \beta} = \frac{1}{2} \left( \frac{\partial v^\alpha}{\partial x^\beta} - \frac{\partial v^\beta}{\partial x^\alpha}
\right).
\end{equation}
Together, eqs.~(\ref{eq:stresstensor_evolution}) and (\ref{eq:rotation_terms}) determine the change of deviatoric
stress due to deformation of the solid body.
\subsection{Equation of state}
The equation of state relates the thermodynamic variables density $\varrho$, pressure $p$, and specific internal energy
$u$. We provide a short description about the equation of states that are implemented in the code.
\subsubsection{Liquid equation of state}
For small compressions and expansions of a liquid, the pressure depends linearly on the change of density.
Assume $\varrho_0$ is the initial density of the uncompressed liquid. Then, the liquid equation of state reads
\begin{equation}
p = c_\mathrm{s}^2 ( \varrho - \varrho_0 ). 
\end{equation}
Here, $c_\mathrm{s}$ denotes the bulk sound speed of the liquid, which is related to the bulk modulus $K_0$ and the initial
density by
\begin{equation}
c_\mathrm{s}^2 = \frac{K_0}{\varrho_0}.
\end{equation}
If the density $\varrho$ falls strongly beyond the value of the initial density $\varrho_0$, the liquid equation of
state fails because the liquid forms droplets and attracting forces vanish. To account for this
behavior, the pressure is set to zero as
soon as the ratio $\varrho/ \varrho_0$ drops below $0.8-0.95$ \citep{melosh1996impact}.
\subsubsection{Murnaghan equation of state}
The Murnaghan equation of state is an extension of the liquid equation of state (see, e.g.,\ \citealt{melosh1996impact}). In contrast to the liquid equation of
state, the pressure now depends nonlinearly on the density
\begin{equation}
p = \frac{K_0}{n_M} \left[ \left( \frac{\varrho}{\varrho_0}\right)^{n_M} -1  \right],
\end{equation}
with the zero pressure bulk modulus $K_0$ and the constant $n_M$. The Murnaghan equation of state is limited to
isothermal compression. Parameters for various materials can be found in the literature, e.g.,\ \cite{melosh1996impact}.
\subsubsection{Tillotson equation of state}
\label{section:tillotson}
The Tillotson equation of state was originally derived for high-velocity impact simulations \citep{til62}.
There are two domains depending upon the material specific energy density $u$. In the case of compressed regions ($\varrho \geq
\varrho_0$) and $u$ lower than the energy of incipient vaporization $E_\mathrm{iv}$ the equation of state reads
\begin{align}\label{eq:tillo1}
p = \left[ a_T + \frac{b_T}{1+u/(E_0 \eta^2)} \right]\varrho u + A_T\chi + B_T\chi^2,
\end{align}
with $\eta = \varrho / \varrho_0$ and $\chi = \eta-1$. 
In case of expanded material ($u$ greater than the energy of complete vaporization $E_\mathrm{cv}$) the
equation of state takes the form
\begin{align}\label{eq:tillo2}
p = & a_T\varrho u +  \left[ \frac{b_T\varrho u}{1+u/(E_0 \eta^2)} \right. \nonumber \\
 & \; + \left. A_T\chi \exp \left\{-\beta_T \left(\frac{\varrho_0}{\varrho}-1\right)\right\} \right] \nonumber \\ & \qquad \qquad
 \times  \exp
 \left\{-\alpha_T\left(\frac{\varrho_0}{\varrho}-1\right)^2\right\}.
\end{align}
The symbols $\varrho_0$, $A_T$, $B_T$, $E_0$, $a_T$, $b_T$, $\alpha_T$, and $\beta_T$ are material dependent
parameters. 

In the partial vaporization $E_\mathrm{iv} < u < E_\mathrm{cv}$, $p$ is linearly interpolated between the pressures
obtained via (\ref{eq:tillo1}) and (\ref{eq:tillo2}), respectively. For a more detailed description see
\cite{melosh1996impact}.

The Tillotson equation of state has the advantage of being computational very simple, while sophisticated enough for the
application over a wide regime of physical conditions, such as high-velocity collisions and impact cratering.

\subsection{Plastic material equations}
Together with the equation of state, the set of equations
(\ref{eq:conservation_of_mass};\ref{eq:conservation_of_momentum};\ref{eq:conservation_internal_energy}) can be used to
describe the dynamics of a solid body in the elastic regime. We apply the von Mises yield
criterion \citep{vonmises1913} to model plasticity. The deviatoric stress is
limited by 
\begin{equation}
S^{\alpha \beta} \rightarrow f_Y S^{\alpha \beta},
\end{equation}
where the factor $f_Y$ is computed from
\begin{equation}
f_Y = \mathrm{min} \left[ Y_0^2/3J_2, 1 \right],
\end{equation}
with the second invariant of the deviatoric stress tensor $J_2$ given by
\begin{equation}
J_2 = \frac{1}{2} S^{\alpha \beta} S_{\alpha \beta},
\end{equation}
and the material dependent yield stress is $Y_0$.
\subsection{Damage model for brittle material}
Since one major application of the newly developed code includes the simulation of the collision between
brittle planetesimals, we included a fragmentation model. Physically, fracture is related to the
failure of atomic or molecular bonds. If an applied force on a solid brittle body is large enough to break the atomic
bonds, the material begins to fracture and develop cracks. A successful continuum model for fragmentation was
derived by \cite{grady:1980} and first implemented in a SPH code by \cite{benz:1995}. The model is based on
two assumptions: brittle materials contain so-called flaws that are sources of weakness leading to activation and
growth under tensile loading, and dynamic fracture depends on the rate of tensile loading. In other words, material that
is slowly pulled apart develops fewer cracks than material that experiences higher strain rates, since one single
crack cannot relieve the higher tensile stress, and hence more flaws get activated leading to more cracks and
consequently more damage.

The scalar parameter damage $d$ parametrizes the influence of cracks on a given volume
\begin{equation}
0 \leq d \leq 1,
\end{equation}
and is defined in a way that $d=0$ represents an undamaged, intact material and $d=1$ a fully damaged material that
cannot undergo any tension or deviatoric stress. In this manner, a material with $d=0.5$ experiences half the stress and
tension of undamaged material with $d=0$.

Damage reduces the strength under tensile loading and the elastic stress $\sigma^{\alpha \beta}$ is decreased
by the factor $(1-d)$, or in more detail
\begin{equation}
\sigma^{\alpha \beta}_d = -\hat{p} \delta^{\alpha \beta} + (1-d) S^{\alpha \beta},
\end{equation}
with
\begin{equation}
\hat{p} = \left\{ \begin{array}{ll} p &  \mathrm{for} \ p \geq 0 \\
  (1-d) p &   \mathrm{for} \ p < 0 \end{array} \right. .
\end{equation}
The number of flaws per unit volume in a brittle body with failure strains that are lower than $\varepsilon$, is given by
the Weibull distribution \citep{weibull:1939}
\begin{equation}
n(\varepsilon) = k \varepsilon^m,
\end{equation}
with the two Weibull parameters $k$ and $m$. Typical values for basalt are $m=16$, $k=\SI{1e61}{\per
\cubic\metre}$ \citep{2007JGRE..112.2001N}.

As soon as a spherical volume $V= \nicefrac{4}{3} \, \pi R_s^3$ undergoes a strain greater than its lowest activation threshold
strain, damage starts to grow according to 
\begin{equation}
\label{eq:damage_growth}
\frac{\mathrm{d}}{\mathrm{d}t} d^{\nicefrac{1}{3}} = \frac{c_\mathrm{g}}{R_s},
\end{equation}
with the constant crack growth velocity $c_\mathrm{g}$. The crack growth velocity is chosen to be 40\% of the velocity of
a longitudinal elastic wave in the damaged material
\begin{equation}
c_\mathrm{g} = \num{0.4} \frac{1}{\varrho} \sqrt{ K_0 + \nicefrac{4}{3} (1-d) \mu}.
\end{equation}
The fracture model requires the distribution of activation threshold strains over the brittle body as a preprocessing
step to the simulations. We show how we achieve this in sect.~\ref{sect:damage_model}.
\section{Numerical model and implementation\label{section:numerical_model}}
In this section, we describe the numerical models that are implemented in the CUDA code. 
\subsection{Smooth particle hydrodynamics}
Smooth particle hydrodynamics is a meshless Lagrangian particle method that was first introduced by
\cite{lucy:1977} and \cite{1977MNRAS.181..375G} for the solution of hydrodynamic equations for compressible flows in
astrophysical applications. The SPH method was extended to model solid bodies at the Los Alamos National Laboratory. This
work was pioneered by \cite{libersky:1991} with various improvements from the same source later on \citep{randles:1996,
libersky:1997}. The first astrophysical application of SPH with strength of material was by \cite{1994Icar..107...98B}.
Our implementation of the SPH equations in the code mainly follows their work.

For a comprehensive introduction to the basic SPH idea and algorithm, we refer to the excellent publications by
\cite{1990nmns.work..269B} and \cite{1992ARA&A..30..543M}. In this section, we present the SPH equations that we have
implemented in the CUDA code, since more than one equivalent SPH representation exists for the partial differential
equations considered in this work.

In the following, roman indices $a$, $b$ denote particle indices and the sum in a SPH equation runs over all of the 
interaction partners.
\subsubsection{Kernel function}
The standard kernel, which is widely used in the astrophysical SPH community and in our CUDA code, is the cubic B-spline
function introduced by \cite{Monaghan85a}. This function is written as
\begin{equation}
W(r;h) = \frac{s}{h^D} 
\left\{    
\begin{array}{l}  
\left(6(r/h)^3 - 6(r/h)^2 +1 \right) \\ \qquad \qquad \mathrm{for\ } 0 \leq r/h < 1/2 \\ \\ 
2\left(1-r/h\right)^3 \\ \qquad \qquad \mathrm{for\ } 1/2 \leq r/h \leq 1 \\ \\
0   \\ \qquad \qquad   \mathrm{for\ } r/h > 1,
\end{array}
\right. 
\end{equation}
where $r$ is the distance between two interacting particles, $h$ denotes the smoothing length, and $s$ is a normalization constant depending on the dimension $D$
\begin{equation}
s = \left\{ \begin{array}{ll} 4/3 & \mathrm{for\ }  1D \\ \frac{40}{7\pi} & \mathrm{for\ }  2D \\ 8/\pi &
\mathrm{for\ }  3D. \end{array} \right.
\end{equation}
The first derivative of the kernel is given by
\begin{equation}
\frac{\partial W(r;h)}{\partial r} = \frac{6s}{h^{D+1}} \left\{ 
\begin{array}{l}
3(r/h)^2 - 2(r/h) \\ \qquad \qquad \mathrm{for\ } 0 \leq r/h < 1/2 \\ \\ 
-(1-r/h)^2  \\ \qquad \qquad \mathrm{for\ } 1/2 \leq r/h \leq 1 \\ \\
0   \\ \qquad \qquad   \mathrm{for\ } r/h > 1.
\end{array}
\right.
\end{equation}
\subsubsection{SPH equations}
In most astrophysical applications of SPH to hydrodynamical problems, 
the continuity equation (\ref{eq:conservation_of_mass}) is not solved. The density of one particle $a$
is given directly by the kernel sum 
\begin{equation}
\label{eq:density_by_kernel_sum}
\varrho_a = \sum_b m_b W_{ab}.
\end{equation}
We use eq.~(\ref{eq:density_by_kernel_sum}) for purely hydrodynamical problems as well and to calculate the densities during the
gravitational collapse of a molecular cloud in sect.~\ref{sect:grav_collapse}.

To model solid bodies, however, it is often more convenient to integrate the continuity equation using the following SPH
representation:
the change of the density of particle $a$ is given by \citep{gray:2001}
\begin{equation}
\label{eq:continuity_equation_sph}
\frac{\mathrm{d} \varrho_a}{\mathrm{d} t} =  \varrho_a \sum_b \frac{m_b}{\varrho_b} (v^\alpha_a - v^\alpha_b)
\frac{\partial W_{ab}}{\partial x_\alpha}.
\end{equation}
In order to calculate the acceleration for particle $a$, we use the 
SPH representation of the equation of motion (\ref{eq:conservation_of_momentum})
\begin{equation}
\label{eq:accel}
\frac{\mathrm{d} v^\alpha_a}{\mathrm{d}t}  = \sum_b m_b \left[ \frac{ \sigma^{\alpha \beta}_a}{\varrho_a^2}
+ \frac{ \sigma^{\alpha \beta}_b}{\varrho_b^2} \right] \frac{\partial W_{ab}}{\partial x^\beta_b}.
\end{equation}
We add some additional artificial terms to eq.~(\ref{eq:accel}) in the following sections  to avoid some numerical
issues of the algorithm.

Finally, the equation for the time evolution of the specific internal energy reads in SPH representation
\begin{equation}
\label{eq:energy_sph}
\frac{\mathrm{d} u_a}{\mathrm{d}t} = \frac{1}{2} \sum_b m_b \left[  \frac{ \sigma^{\alpha \beta}_a}{\varrho_a^2}
+ \frac{ \sigma^{\alpha \beta}_b}{\varrho_b^2} \right] 
\left( v_a^\alpha - v_b^\alpha \right)  \frac{\partial W_{ab}}{\partial x^\beta_b}.
\end{equation}
\subsubsection{Artificial viscosity}
An important issue is the artificial viscosity. In order to
treat shock waves in an ideal fluid, dissipation is needed to dissipate kinetic energy in the shock front. In SPH, it is
necessary to prevent the particles from unphysical mutual penetration. The artificial viscosity terms were introduced by 
\cite{monaghan:1983}. They can be regarded as an additional artificial pressure term in the equations for the
conservation of momentum and energy. The additional pressure is calculated for each interaction pair $a$ and $b$ as
follows: 
\begin{equation}
\Pi_{ab} = \frac{-\alpha {\overline{c}_\mathrm{s}}_ab  \nu_{ab} + \beta \nu_{ab}^2}{\overline{\varrho}_{ab}},
\end{equation}
where $\alpha$ and $\beta$ are free parameters that determine the strength of the viscosity, and
$\overline{\varrho}_{ab}$ and ${\overline{c}_\mathrm{s}}_{ab}$ are the averaged quantities for density and sound speed for
the two interacting particles, $\overline{\varrho}_{ab} = (\varrho_a + \varrho_b) / 2 $ and
${\overline{c}_\mathrm{s}}_{ab} = ({c_\mathrm{s}}_a + {c_\mathrm{s}}_b ) / 2$.

The term $\nu_{ab}$ is an approximation for the divergence and is calculated accordingly
\begin{equation}
\nu_{ab} = \frac{\overline{h}_{ab} (\vect{v_a} - \vect{v_b} ) \cdot ( \vect{x_a} - \vect{x_b})}{(\vect{x_a} -
\vect{x_b})^2 + \varepsilon_v \overline{h}_{ab}^2}.
\end{equation}
Here, $\overline{h}_{ab}^2$ is the average of the smoothing length of particle $a$ and $b$.
The term $\varepsilon_v \overline{h}_{ab}^2$ in the denominator  prevents divergence of $\nu_{ab}$ for particles with
small separations.

The artificial pressure is added to the acceleration of the particle
\begin{equation}
\label{eq:accel_artvisc}
\frac{\mathrm{d} v^\alpha_a}{\mathrm{d}t}  = \sum_b m_b \left[ \frac{ \sigma^{\alpha \beta}_a}{\varrho_a^2}
+ \frac{ \sigma^{\alpha \beta}_b}{\varrho_b^2} + \Pi^\star_{ab} \right ] \frac{\partial W_{ab}}{\partial x^\beta_b},
\end{equation}
with $\Pi^\star_{ab} = \Pi_{ab}$ for $(\vect{v_a} - \vect{v_b} ) \cdot ( \vect{x_a} - \vect{x_b}) < 0$ and
$0$ elsewhere. This means, the artificial viscosity leads to a repellent force only for approaching
particles.
\subsubsection{Artificial stress: Tensile instability}
The tensile instability can occur in SPH simulations that include materials with equation of states that allow negative
pressure or tension. In the worst case, the instability leads to artificial numerical clumping of particles and in
the end to unnatural fragmentation.
There are several different approaches to overcome the tensile instability. In our code, we have implemented an
additional artificial stress.
To avoid artificial clumping of particles due to the tensile instability of the SPH method, \cite{monaghan:2000} added a
small repulsive artificial stress to the SPH equation (\ref{eq:accel}) 
\begin{equation} \frac{\mathrm{d}
v^\alpha_a}{\mathrm{d}t}  = \sum_b m_b \left[ \frac{ \sigma^{\alpha \beta}_a}{\varrho_a^2} + \frac{ \sigma^{\alpha
\beta}_b}{\varrho_b^2} + \zeta^{\alpha \beta}_{ab} f^n+  \Pi^\star_{ab} \right ] \frac{\partial W_{ab}}{\partial x^\beta_b}.
\end{equation}
The artificial stress $\zeta^{\alpha\beta}_{ab}$ is given by the sum of the artificial stress for particle $a$ and $b$,
respectively,
\begin{equation}
\zeta^{\alpha \beta}_{ab} = \zeta^{\alpha \beta}_a + \zeta^{\alpha \beta}_b,
\end{equation}
with $\zeta^{\alpha \beta}_a$ for tension
\begin{equation}
\zeta^{\alpha \beta}_a = -\varepsilon_s \frac{\sigma^{\alpha \beta}_a}{\varrho^2_a},
\end{equation}
whereas $\zeta^{\alpha \beta}_a = 0$ for compression. The factor $\varepsilon_s$ is typically around $0.2$. The repulsive
force that results from the artificial stress is scaled with $f^n$, with $n>0$, where $f$ is given by the ratio between
the kernel for the interaction of particles $a$ and $b$ and $\Delta mpd$, the mean particle distance in the vicinity of
particle $a$, 
\begin{equation}
f = \frac{W(r_{ab})}{W(\Delta mpd)}.
\end{equation}
The usual approach is to take the mean particle distance of the initial particle distribution for $\Delta
mpd$.
The artificial stress leads to an additional repulsive force whenever tension would trigger the tensile instability. 
The effect of artificial stress is shown in an illustrative example in sect.~\ref{section:rings}.
\subsubsection{XSPH --- The Velocity}
The time derivative of the location of particle $a$ is given by the velocity of particle $a$
\begin{equation}
\frac{\mathrm{d} \vect{x_a}}{\mathrm{d} t} = \vect{v_a}.
\end{equation}
However, the velocity of the continuum cannot necessarily be identified with the particle velocity. 
The XSPH algorithm \citep{monaghan:1989} moves the particles at an averaged speed smoothed by the velocity of the
interaction partners
\begin{equation}
\frac{\mathrm{d} \vect{x_a}}{\mathrm{d} t} = \vect{v_a} + x_{\mathrm{sph}} \sum_b \frac{2m_b}{\varrho_a \varrho_b}
\left(\vect{v_b} - \vect{v_a} \right) W_{ab}. 
\end{equation}
Here, the factor $x_{\mathrm{sph}}$ is called the XSPH factor, $0 \leq x_{\mathrm{sph}} \leq 1$. Throughout our
simulations performed for this paper, we set $x_{\mathrm{sph}}=0.5$ when we use XSPH.
\subsubsection{A note on linear consistency}
In standard SPH, the velocity derivatives in
eq.~(\ref{eq:strain_rate_tensor}) for the determination of the strain
rate tensor for particle $a$ are given by
\begin{equation}
\frac{\partial v_a^\alpha}{\partial x_a^\beta} = \sum_b
\frac{m_b}{\varrho_b} (v^\alpha_b - v^\alpha_a) \frac{\partial
W_{ab}}{\partial x^\beta_a}.
\end{equation}
This approach, however, leads to erroneous results and does not conserve
angular momentum due to the discretization error by particle disorder in
simulations of solid bodies. This error can be avoided by constructing
and applying a correction tensor \citep{2007A&A...470..733S}
\begin{equation}
\frac{\partial v_a^\alpha}{\partial x_a^\beta} = \sum_b
\frac{m_b}{\varrho_b} (v^\alpha_b - v^\alpha_a) \sum_\gamma \frac{\partial
W_{ab}}{\partial x^\gamma_a}C^{\gamma \beta},
\label{eq:ext_sph}
\end{equation}
where the correction tensor $C^{\gamma \beta}$ is given by the inverse of
\begin{equation}
\sum_b \frac{m_b}{\varrho_b} (x^\alpha_a - x^\alpha_b) \frac{\partial
W_{ab}}{\partial x^\gamma_a},
\end{equation}
and it holds
\begin{equation}
\sum_b \frac{m_b}{\varrho_b} (x^\alpha_b - x^\alpha_a) \sum_\gamma
\frac{\partial W_{ab}}{\partial x^\gamma_a} C^{\gamma \beta} = \delta^{\alpha
\beta}.
\end{equation}
We apply the correction tensor according 
to eq.~(\ref{eq:ext_sph}) in all our simulations that demand for derivatives of the velocity to force first order consistency.
\subsection{Damage model \label{sect:damage_model}} 
As a preprocessing step to a simulation containing brittle materials, we have to distribute activation threshold strains
or flaws among the SPH particles. We apply the algorithm introduced by \cite{benz:1995} with the following approach: Flaws
are distributed randomly among the brittle particles until each particle has been assigned with at least one activation
threshold. We let $N_\mathrm{flaw}$ be the number of totally distributed activation flaws. For each flaw $j$, $1\leq j \leq
N_\mathrm{flaw}$, a particle $a$ is chosen randomly and assigned with the activation threshold strain derived from the
Weibull distribution $\varepsilon_{a,j}$,
\begin{equation}
\varepsilon_{a,j} = \left[ \frac{j}{kV} \right]^{\nicefrac{1}{m}},
\end{equation}
where $V$ is the volume of the brittle material and $k$ and $m$ are the material dependent Weibull parameters. Using
this procedure, each particle gets at least one flaw and on average a total of $N_\mathrm{flaws} = N_p \ln N_p$ flaws
are distributed. This number of flaws has to be stored in the memory during the simulation and is the
main memory consumer in simulations with brittle materials.

During the simulation, once a particle undergoes a strain greater than its lowest activation threshold strain, damage
grows according to a modified version of eq.~(\ref{eq:damage_growth})
\begin{equation}
\frac{\mathrm{d}}{\mathrm{d} t} d^{\nicefrac{1}{3}} = n_\mathrm{act} \frac{c_\mathrm{g}}{R_s},
\end{equation}
where $n_\mathrm{act}$ is the number of activated flaws of the particle. This modification accounts for the accumulation
of cracks.

Additionally, a damage limiter is included in the model. Since a particle has in general more than only one activation
threshold, its behavior under tensile loading would be given by the flaw with the lowest activation threshold.  To
prevent this, the damage of the particle is limited by the ratio between the number of activated flaws $n_\mathrm{act}$
and the number of total flaws $n_\mathrm{tot}$ of this particle
\begin{equation}
d_\mathrm{max} = \frac{n_\mathrm{act}}{n_\mathrm{tot}}.
\end{equation}
In other words, a particle can only be totally damaged if all of its flaws are activated.

In order to determine if a particle exceeds one of its threshold strains during the simulation, we need to derive the
local scalar strain from the three-dimensional stress. \cite{benz:1995} suggest computing the local scalar strain
$\varepsilon$ from the maximum tensile stress $\sigma_\mathrm{max} = \max [\sigma_1, \sigma_2, \sigma_3 ]$, where
$\sigma_\gamma$ are the principal stresses determined by a principal axis transformation of the stress tensor
$\sigma^{\alpha \beta}$, which is already reduced by damage and yield strength
\begin{equation}
\varepsilon = \frac{\sigma_\mathrm{max}}{(1-d) E}.
\end{equation}
The Young's modulus $E$ of the intact material given by the bulk and shear moduli is written as
\begin{equation}
E = \frac{9 K \mu}{3K+\mu}.
\end{equation}
\subsection{Barnes-Hut tree} 
We have implemented the efficient routines for Barnes-Hut tree builds and calculation of the gravitational forces for GPUs introduced by \cite{burtscher:2011}. The algorithm
for the tree generation is explained in sect.~\ref{sect:cuda_tree}.

\cite{1986Natur.324..446B} were the first to use a hierarchical tree to calculate the gravitational forces between N
bodies. Their approximation algorithm uses a tree-structured hierarchical subdivision of space into cubic cells, each of
which is divided into $2^D$ subcells. The algorithm has the order ${\cal O}(N \log N)$ compared to the ${\cal O}(N^2)$
algorithm of the direct sum. The basic idea is that only nearby particles have to be treated individually during the
calculation of the gravitational force on a specific particle. Distant nodes of the tree can be seen as one single
particle sitting at the center of mass of the specific tree node and carrying the mass of all the particles that reside
in this node. Although different algorithms for the computation of self-gravitational forces exist, the hierarchical
tree method is specially suited for the use in combination with SPH. As a side product of the computation of the
gravitational forces, the hierarchical tree is constructed for each individual time step and hence, a geometrical hierarchy
of the SPH particles (leaves of the tree) is available at all time of the computation. It can be used to determine the
interaction partners for the SPH algorithm by performing a tree walk for each individual particle \citep{hernquist:1989}. Although the tree
walks may take longer than the search through a helping grid, the interaction partner search may be faster because, in
any case the tree build needs to be carried out in simulations with self-gravity. 

We calculate the center of mass and total mass of every node in the tree to calculate the gravitational force exerted on the SPH particles. We call these objects pseudo-particles in the following. Now, for each SPH
particle, we perform a tree walk over all nodes.
With the use of the
$\vartheta$ criterion, we identify whether a pseudo-particle sitting in the node may be used for the calculation of the
gravitational force acting on our particle or whether we have to descend further down in the tree. The $\vartheta$ criterion is given by
\begin{equation}
\frac{s_{d_t}}{r_{pp}} < \vartheta,
\end{equation}
where $s_{d_t}$ denotes the edge length of the tree node containing the pseudo-particle and $r_{pp}$ is the distance
between the SPH particle and the pseudo-particle.
The gravitational force from the pseudo-particle acting on the particle $a$ is then given by
\begin{equation}
\label{eq:gravitational_force}
\vect{F_a} = \frac{Gm_a m_p}{r_{pp}^3} \left( \vect{x_a} - \vect{x_p} \right),
\end{equation}
where $\vect{x_p}$ and $m_p$ denote the location and the mass of the pseudo-particle, respectively.
To avoid numerical instabilities, the distance between the pseudo-particle and the particle is smoothed.

\cite{1989ApJS...70..389B} present a detailed error analysis for their tree algorithm. They find consistent accuracies
of the algorithm for $\vartheta=0.3$  with the consideration of the monopoles of the tree compared to $\vartheta=0.7$
and the additional calculation of the quadrupoles for each tree node. The error scales with $\vartheta^2$ and $\sqrt{N_p}$. For the sake of memory saving, we calculate only the monopoles of the tree and use
$\vartheta \leq 0.5$ in our simulations including self-gravity instead of the usual value of $0.8$.
\subsection{Time integration}
An important subset of every SPH code is the time integration function. We have implemented two different time
integrators for the code: a predictor-corrector algorithm following the publication by \cite{monaghan:1989} and a
Runge-Kutta 3rd order integrator with adaptive time step control. In the following, we present the basic algorithm
for the integrators and the time step control mechanisms.

\subsubsection{Predictor-corrector}
A simple predictor-corrector integration scheme is implemented in the code as follows. We let $q_a(t)$ be a quantity that is
integrated in time and $\frac{\mathrm{d}}{\mathrm{d}t} q_a$ its derivative. At first we perform a prediction integration
step and calculate the value after half the time step $\Delta t$,
\begin{equation}
q_a^{(p)} = q_a  + \frac{\Delta t}{2} \frac{\mathrm{d}}{\mathrm{d}t} q_a.
\end{equation}
Now, we use the predicted values $q_a^{(p)}$ to calculate the time derivatives for the correction step and based on them the 
new values $q_a(t+\Delta t)$,
\begin{equation}
q_a^{(c)} = q_a + \Delta t \frac{\mathrm{d}}{\mathrm{d}t} q_a^{(p)}.
\end{equation}
Obviously, the time step $\Delta t$ has to be limited. To set an upper limit to the time step we use the following
condition from the acceleration:
\begin{equation}
\Delta t _\mathrm{max,1} = \min_{a \in N_p} \left[ 0.2 \times \frac{h_a}{\left| \frac{\mathrm{d} \vect{v_a}}{\mathrm{d}t}
\right| } \right] 
\end{equation}
and the Courant condition
\begin{equation}
\Delta t_\mathrm{max,2} = \min_{a \in N_p} \left[  0.7 \times \frac{h_a}{c_\mathrm{s_a}} \right].
\end{equation}
The maximum time step for this integration step is then given by 
\begin{equation}
\Delta t _\mathrm{max} = \min \left[ \Delta t_\mathrm{max,1}  , \Delta t_\mathrm{max,2} \right] .
\end{equation}
Additionally, we allow only a maximum change for all other quantities that are integrated. This may be necessary for
quantities, such as the density, stress, internal energy, and damage, and is written as
\begin{equation}
\Delta t _\mathrm{max} = \min_{a \in N_p} \left[ 0.7 \times \sqrt{
\frac{\left[q_a^{(p)}\right]^2}{\left[\frac{\mathrm{d}}{\mathrm{d}t} q_a\right]^2}} \right].
\end{equation}
\subsubsection{Embedded Runge-Kutta integrator}
In addition to the predictor-corrector scheme, we  also implemented a more sophisticated Runge-Kutta integration
scheme that features an adaptive step size control. The basic algorithm is based on a Runge-Kutta 3rd order scheme
where the third step that is used to give an error estimate about the quality of integration step; see \cite{abramowitz1964handbook} and especially \cite{butcher:1987} for a more detailed description. The
great advantage of this scheme lies in the possibility of setting a maximum relative error for the quantities since the
Courant condition, in principle, is only applicable in hydrodynamical simulations.

The standard 3rd order Runge-Kutta integration scheme for a quantity $q$, which is integrated in time from
$t_n$ to $t_n + \Delta t$, is written as
\begin{equation}
q_{n+1}^{\mathrm{(rk3)}} = q_n + \Delta t \left[ \frac{1}{6} k_1 + \frac{4}{6} k_2 + \frac{1}{6} k_3 \right]
\end{equation}
with $q_n=q(t_n)$ and $q_{n+1} = q(t_n + \Delta t)$.

Writing $f(t_n,q_n) = \frac{\mathrm{d} q_n}{\mathrm{d}t}$ the three substeps $k_1,k_2,k_3$ are given by
\begin{align}
k_1 &= f(t_n,q_n) \\
k_2 &= f\left(t_n + \frac{\Delta t}{2} , q_n + \frac{\Delta t}{2} k_1\right) \\
k_3 &=  f(t_n + \Delta t, q_n - \Delta t k_1 + 2 \Delta t k_2).
\end{align}
The quantity can also be integrated using the Euler midpoint method. There, the value at time $t_n+\Delta t$ is given by
\begin{equation}
q_{n+1}^{\mathrm{(mm)}} = q_n + \Delta t f\left(t_n+\frac{\Delta t}{2}, q_n+\frac{\Delta t}{2} f(t_n, y_n)\right).
\end{equation}
Now, we use the difference between $q_{n+1}^{\mathrm{(mm)}}$ and $q_{n+1}^{\mathrm{(rk3)}}$ to give an estimate for
the local error of the scheme by comparison to the value resulting from an Euler step
\begin{equation}
\varepsilon_{\mathrm{err}} = \frac{\left| q_{n+1}^{\mathrm{(mm)}} - q_{n+1}^{\mathrm{(rk3)}}\right|}{\left|q_n + \Delta t f(t_n,q_n)\right|}.
\end{equation}
If $\varepsilon_{\mathrm{err}}$ is smaller than the desired relative error defined in the configuration file
$\varepsilon_{\mathrm{des}}$, the integration step
is accepted and the time step size is increased to
\begin{equation}
\Delta t _\mathrm{new} = \Delta t  \left|
\frac{\varepsilon_{\mathrm{des}}}{\varepsilon_{\mathrm{err}}}\right|^{0.3}.
\end{equation}
If the error is too large, the time step size is decreased accordingly to
\begin{equation}
\Delta t _\mathrm{new} = 0.9 \Delta t  \left|
\frac{\varepsilon_{\mathrm{des}}}{\varepsilon_{\mathrm{err}}}\right|^{\nicefrac{1}{4}}.
\end{equation}
\subsection{Implementation in CUDA}
In this section, we present how we have implemented the numerical algorithms from the last section for the NVIDIA GPUs
with CUDA. Depending on the problem, the information about the particles is stored in one-dimensional arrays, e.g.,\ {\tt
posx[$N_p$], posy[$N_p$], posz[$N_p$], velx[$N_p$], vely[$N_p$], velz[$N_p$],} and {\tt stress[$D\times D \times N_p$]}. The bottleneck for GPU computation is the slow memory
bandwidth between host and device. Therefore, we initially copy all particle data to the GPU and perform all
calculations with the SPH particles there. After certain time frames the data is written to the hard disk in an
additional POSIX thread to avoid that the computation time depends on disk i/o for simulations with high output.
\subsubsection{Barnes-Hut tree\label{sect:cuda_tree}}
We have implemented the Barnes-Hut tree strictly following the efficient implementation for N-body calculations with
CUDA by \cite{burtscher:2011}. The hierarchical decomposition is recorded in an octree. Since CUDA does not allow the
use of pointers, which is convenient for the handling of trees, e.g.,\ allows for recursive programming of tree walks, we
use one large index array for all SPH particles and all tree nodes, which we call {\tt childList}. This memory is
allocated dynamically at the start of the simulation run. The maximum
number of tree nodes is given by $\mathrm{ceil}[2.5 \times N_p]$, where $N_p$ is the number of SPH particles in the
simulation. For each tree node there are $N_\mathrm{children}$ entries in {\tt childList}, where $N_\mathrm{children}$
is simply given by the dimension
\begin{equation}
N_\mathrm{children} = 2^D.
\end{equation}
The tree is constructed in a way that the $N_p$ particles reside in $N_p$ leaves in the end. 
At first, the root of the tree is constructed from the geometry of the particle distribution. The computational domain
(in three dimensions) is given by the cube, which encloses all SPH particles. This cube has the edge length
\begin{equation}
s_0 = \max_{\alpha \in D}  \left[ \max_{a,b \in N_p} \left[ x_a^\alpha - x_b^\alpha \right] \right]
\end{equation}
and defines the edge length of the root node.
The edge length of a tree node of depth $d_t$ is given by
\begin{equation}
s_{d_t} = \left[ \frac{1}{2}\right]^{d_t} s_0.
\end{equation}
Starting from the root node, the particles are inserted into the tree in the following way. In the beginning, all but
one (the root node) entry in {\tt childList} are $-1$, corresponding to the status "no kid".  All particles are
distributed among the threads on the GPU.  Now $N_\mathrm{threads}$ threads on the GPU start to compare the locations of their
SPH particles with the position of the root node to determine the childindex. Since there are many more threads than
children (especially in the beginning of the algorithm), we need to avoid race conditions actively. To accomplish this, we add the
status $-2$ or "locked" to the {\tt childList} and use the CUDA\-function "atomic compare and save" {\tt atomicCAS} to
lock an entry in the {\tt childList}. This functions checks first if the specific index is already locked. If the index
in {\tt childList} is already locked, the thread waits for the other threads to finish and restarts its attempt. If the
index is not locked, the threads read the index value, which may be $-1$ in the beginning or a different node index,
and saves it. Now, the second round starts and the children of the root node are already filled with SPH particles that
were sorted there. Since leaves or particles cannot have children by definition, new inner tree nodes have to be
generated. A new inner tree node has half of the edge length from its parent. New inner tree nodes are added until both
particles, the new particle and the particle in the parent node, can be sorted into two different leaves. As soon as
both particles reside in different leaves, the thread writes and cancels the lock. 

When we want to calculate the gravitational forces between the SPH particles, we have to calculate the positions
and locations of the pseudo-particles in the tree. We use the algorithm presented by \cite{burtscher:2011} in the
following manner to determine the pseudo-particle in a specific node: For each inner node, we start in parallel for
$N_\mathrm{threads}$ a CUDA kernel that picks the last $N_\mathrm{threads}$ added inner nodes, for which we 
calculate the center of mass and total mass at first. If the inner nodes are not leaves, the thread looking at these
nodes and has to wait for all other threads to end beforce continuing because it needs the location and mass of the
pseudo-particles in these nodes. This is again done by the use of $-1$ as the initial mass of a pseudo-particle. When the
thread finishes the calculation for its node, it moves up one tree depth and continues with the next node. The
implementation of Burtscher is especially designed for CUDA, regarding memory access and parallelization.
\subsubsection{Neighbor search}
One major task in a SPH code is to find an efficient way to determine the interaction partners for each SPH particle for the
kernel sums. Normally, the fastest way to obtain the interaction partner list in a SPH simulation with one fixed smoothing
length is via an additional helping grid with a grid size of the smoothing length. Then all particles are
sorted into the grid, and only the neighboring grid cells have to be searched for interaction partners. There are
numerous publications about the best way to implement a searching grid in SPH, and we refer to the nice overview given
in \cite{speith:1998}.

Since we already computed the Barnes-Hut tree for the calculation of self-gravity, we can easily benefit and use
the tree to determine the interaction partners without any additional helping grid. The tree search is implemented in
the following way: Corresponding to the {\tt childList} for the tree, we also have an index array that stores the
information about the interaction partners, {\tt interactionsList}. In contrast to the tree generation algorithm, there
are no race conditions in the interaction partner search algorithm, since we can search for the interaction partners of
particle $a$ independently from the search of particle $b$. Each thread searches for its particles and starts the search for
each particle at the root node. If the particle lies within the cube of edge length
\begin{equation}
l_{i} = s_{d_t}/2 + h,
\end{equation}
we descend into the children and continue searching. As soon as a leaf, that is a different particle, is encountered,
we check whether the distance between the leaf and particle is smaller than the smoothing length, and if so, the
index of the leaf is added to the {\tt interactionsList} accordingly. 

Since all threads are independent in this algorithm, it is very efficient and fast.

\subsubsection{Smooth particle hydrodynamics equations}
The CUDA implementation of the SPH sums for the equation of motion and conservation of energy is straightforward, since
as soon as the interaction partners for each particle are known, the computation of these quantities can be performed
independently. 

\section{Numerical tests and code validation\label{section:tests}}
In this section, we present some of our numerical tests that we performed for code validation.
We chose three standard cases to test the different aspects of the code.

\subsection{Colliding rubber cylinders{\label{section:rings}}}
A suitable test for the correct implementation of the tensile instability fix is the simulation of two
colliding rubber cylinders. As first presented by \cite{swegle:1992}, the collision results in fragmentation, especially
when the particles are initially placed on a squared grid. We use the setup by \cite{monaghan:2000} and
\cite{gray:2001}, and use following material parameters and units: The unit of density is $\varrho_0$, the unit of
length is $\num{1}~\si{cm}$, the unit of velocity is $c_\mathrm{s}$ ($\num{852}~\si{m/s}$), and the unit of stress is $\varrho_0
c_\mathrm{s}^2$. The value of the shear modulus $\mu$ is set to $\num{0.22}~\varrho_0 c_\mathrm{s}^2$. We use two rubber cylinders with
inner radius $\num{3}~\si{cm}$ and outer radius $\num{4}~\si{cm}$, moving initially in one direction with $\pm
\num{0.059}~c_\mathrm{s}$, so that the collision velocity is $\num{0.118}~c_\mathrm{s}$. 
In order to test the implementation of artificial stress, we perform two different simulations: 
We set the parameters for the
artificial stress to $\varepsilon_s=0.2$ and $n=4$ in one
simulation, while we do not add any artificial stress in the other run.
Additionally, XSPH is applied with $x_\mathrm{sph}=0.5$ and the artificial viscosity coefficient is set to $\alpha = \num{1}$ in
both setups.

\begin{figure}
\includegraphics[width=0.48\textwidth]{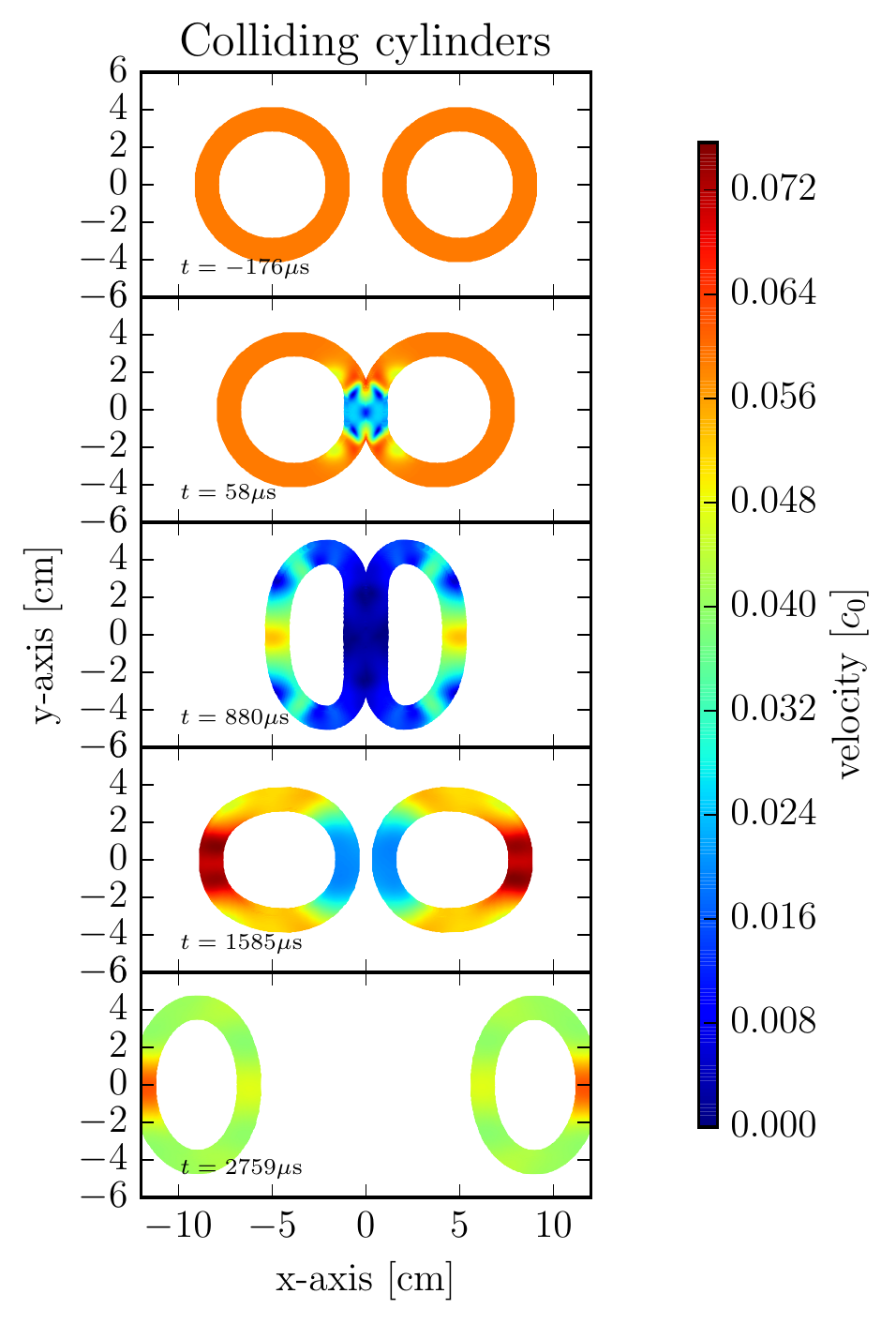}
\caption{Colliding rubber cylinders with artificial stress; absolute value of the velocity is
color-coded.
\label{fig:colliding_rings}
The use of an additional artificial stress hinders the fracture of the colliding rubber cylinders. The moment of first
contact between the two cylinders is at $t=0$.}
\end{figure}

\begin{figure}
\includegraphics[width=0.48\textwidth]{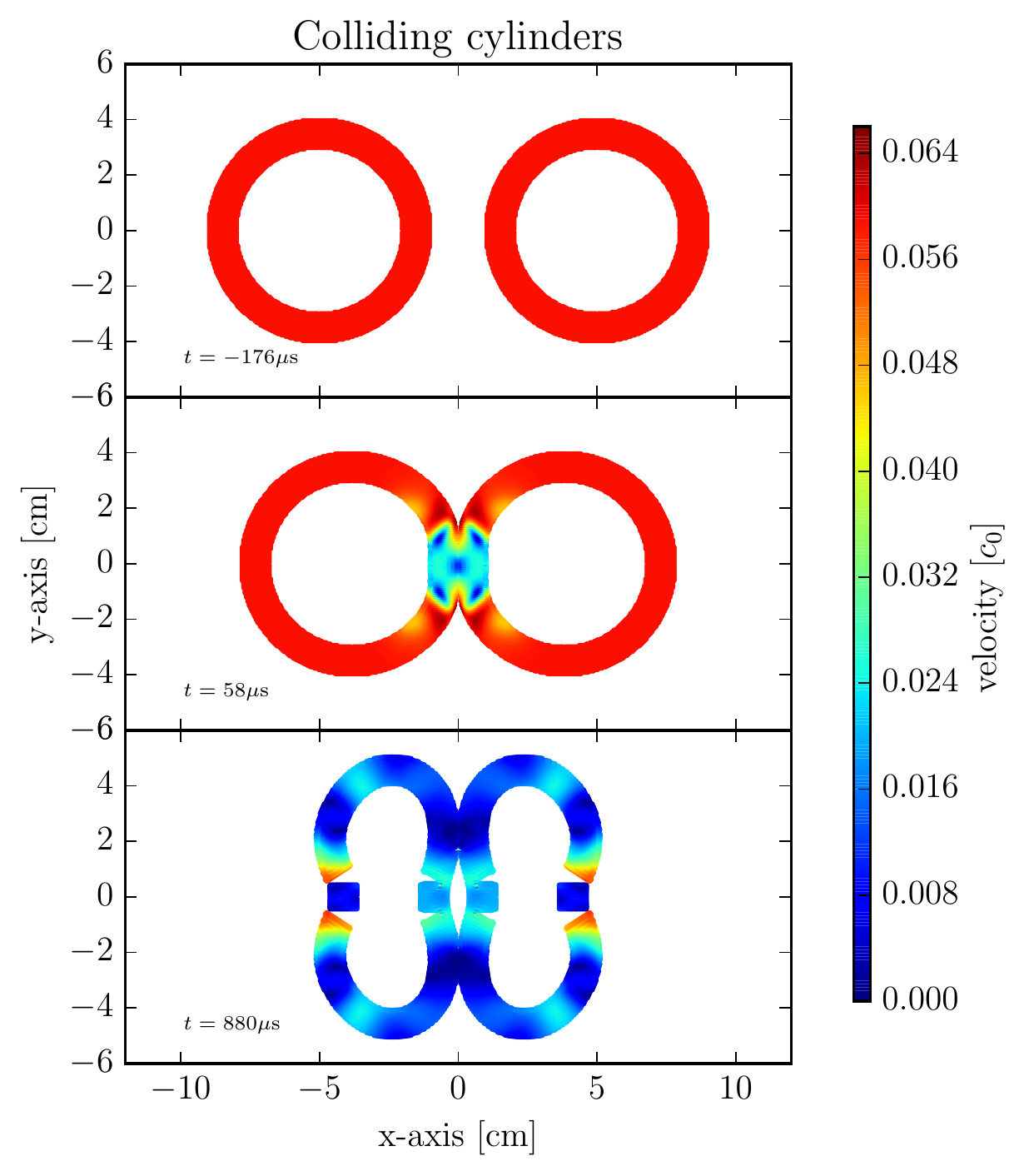}
\caption{Colliding rubber cylinders without artificial stress; absolute value of the velocity is
color-coded.
Without the use of an additional artificial stress, the colliding rubber cylinders break up. The time of first
contact between the two cylinders is $t=0$
\label{fig:colliding_rings_noat}
}
\end{figure}

The results of the simulations with and without artificial stress are shown in fig.~\ref{fig:colliding_rings} and
\ref{fig:colliding_rings_noat}, respectively. In the case without any artificial stress, the cylinders fracture shortly
after maximum compression.  In the simulation with artificial stress, the cylinders bounce off each other without
fracturing and begin to swing. The timescale is chosen according to \cite{monaghan:2000} for best comparison. The
cylinders touch at $t=0$.
Our results are in perfect agreement with \cite{monaghan:2000} and \cite{gray:2001}.
\subsection{Gravitational Collapse of a Molecular Cloud\label{sect:grav_collapse}}
In order to test the gravitational tree algorithm in the code, we run the standard test case for self-gravitating
objects presented by \cite{1979ApJ...234..289B}. These authors investigated the collapse of a self-gravitating, initially
spherical, and uniformly rotating gas cloud. Internal pressure is included, but viscous and
magnetic forces are neglected. The collapse of the cloud is considered to be isothermal. The equation of state is
that of an ideal gas of pure molecular hydrogen with constant temperature $T=\SI{10}{\kelvin}$. The initial
conditions are as follows: the total mass of the molecular cloud is $M=1.0~M_{\odot}$, the mean density is given by
$\varrho_0 = \SI{1.44e-14}{kg\per \cubic \metre}$, the molecular cloud rotates uniformly with the angular velocity
$\Omega = \SI{1.6e-12}{rad/s}$, and the initial radius of the spherical cloud is $R_0=\SI{3.2e14}{m}$.

A nonaxisymmetric perturbation of mode $m=2$ and amplitude $0.5$ depending on the azimuthal angle $\varphi$ is
superposed on the initially uniform density
\begin{equation}
\varrho(\varphi) = \varrho_0 \left[ 1 + \frac{1}{2} \cos \left( 2 \varphi \right) \right].
\end{equation}
The free-fall time is given by 
\begin{equation} t_{f} = \left[ \frac{3\pi}{32 G \varrho_0} \right]^{\frac{1}{2}} =
\SI{5.529e11}{s}. 
\end{equation} 
\begin{figure}
\includegraphics[width=0.48\textwidth]{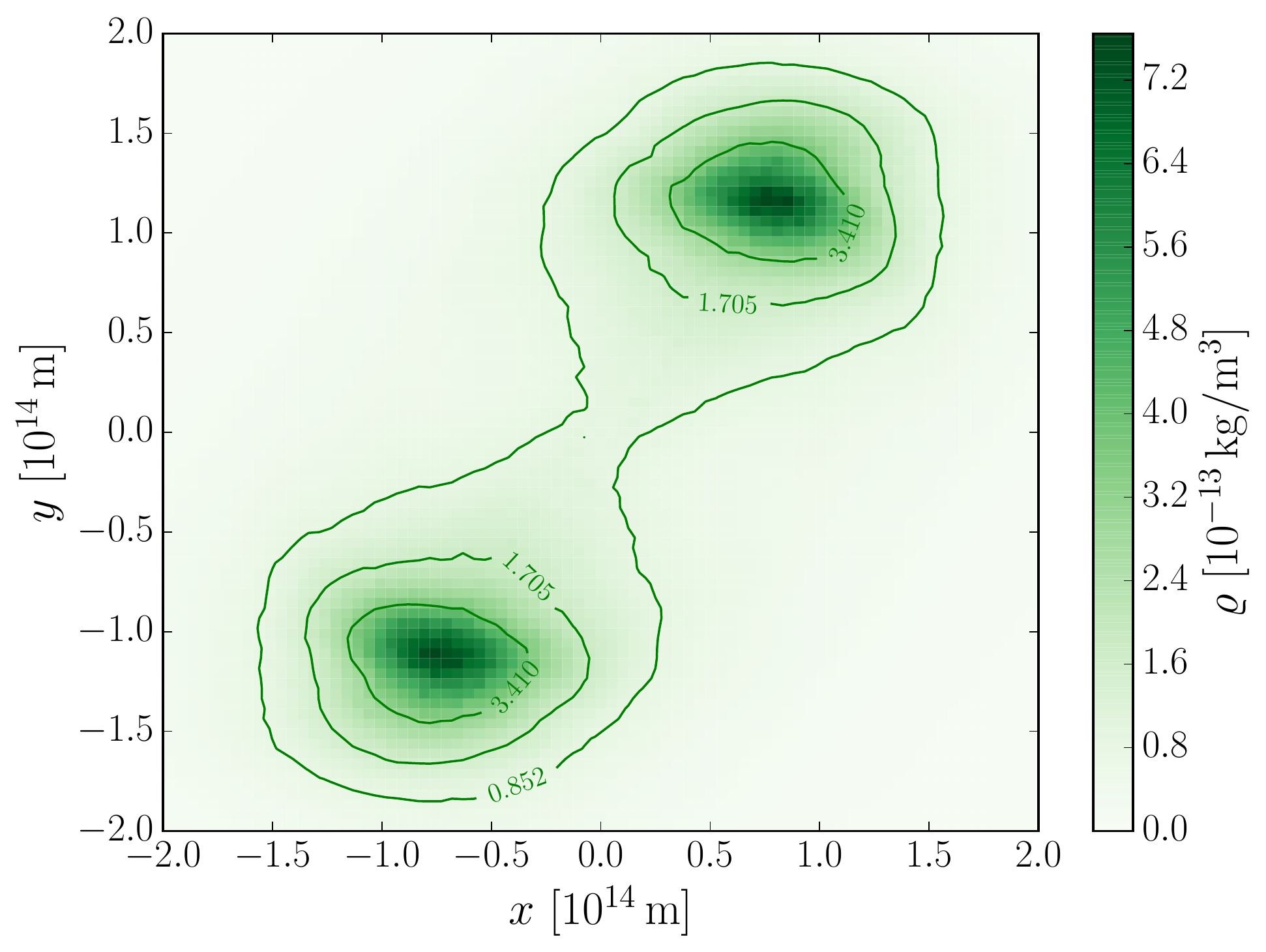}
\caption{Gravitational collapse, value of the density in the midplane is color-coded.
\label{fig:gravitational_collapse}
The density in the midplane for $z=0$ is shown at $t=\SI{0.998}{t_f}$. In order to generate the plot, the $\num{2e6}$ SPH
particles were mapped on a cubic grid $128 \times 128 \times 128$.
}
\end{figure}
We performed several simulations with different number of particles, ranging from $\num{8000}$ to $\num{2000000}$. We
varied the type of smoothing of the gravitational forces. Since the mean particle distance rapidly shrinks during
the simulation, it is convenient to use a variable smoothing length and a constant number of interaction partners. The
kernel value for the interaction of particle $a$ and $b$ is then calculated with the mean smoothing length
\begin{equation}
W_{ab}(r_{ab};h_{ab}) = W_{ab} ( r_{ab}; \nicefrac{1}{2}(h_a + h_b)).
\end{equation}
To enforce a constant number of interaction partners per particle, we start with the smoothing length of the
last time step and iterate using the new smoothing length \citep{hernquist:1989}
\begin{equation}
h_{\mathrm{new}} = h \times \frac{1}{2} \left( 1 + \left[ \frac{n_\mathrm{des}}{n_\mathrm{a}} \right]^{1/D} \right),
\end{equation}
where $n_\mathrm{des}$ is the desired number of interaction partners and $n_\mathrm{a}$ the actual number of
interaction partners gained with the smoothing length $h$.

The results of our simulations are in very good agreement to previous calculations by \cite{1979ApJ...234..289B}.
Moreover, in accord with \cite{bate:1997} we find differing collapse speeds depending on the smoothing of the
gravitational potential. 

Figure~\ref{fig:gravitational_collapse} shows a color-coded plot of the density in the midplane $z=0$ after a
simulation time of approximately the free fall time $t_f$ for the simulation with two million SPH particles.  There is
already an increase of density of almost two orders in magnitude at the location of the initial perturbation, where the
two protostars are about to form.

\begin{figure}
\includegraphics[width=0.48\textwidth]{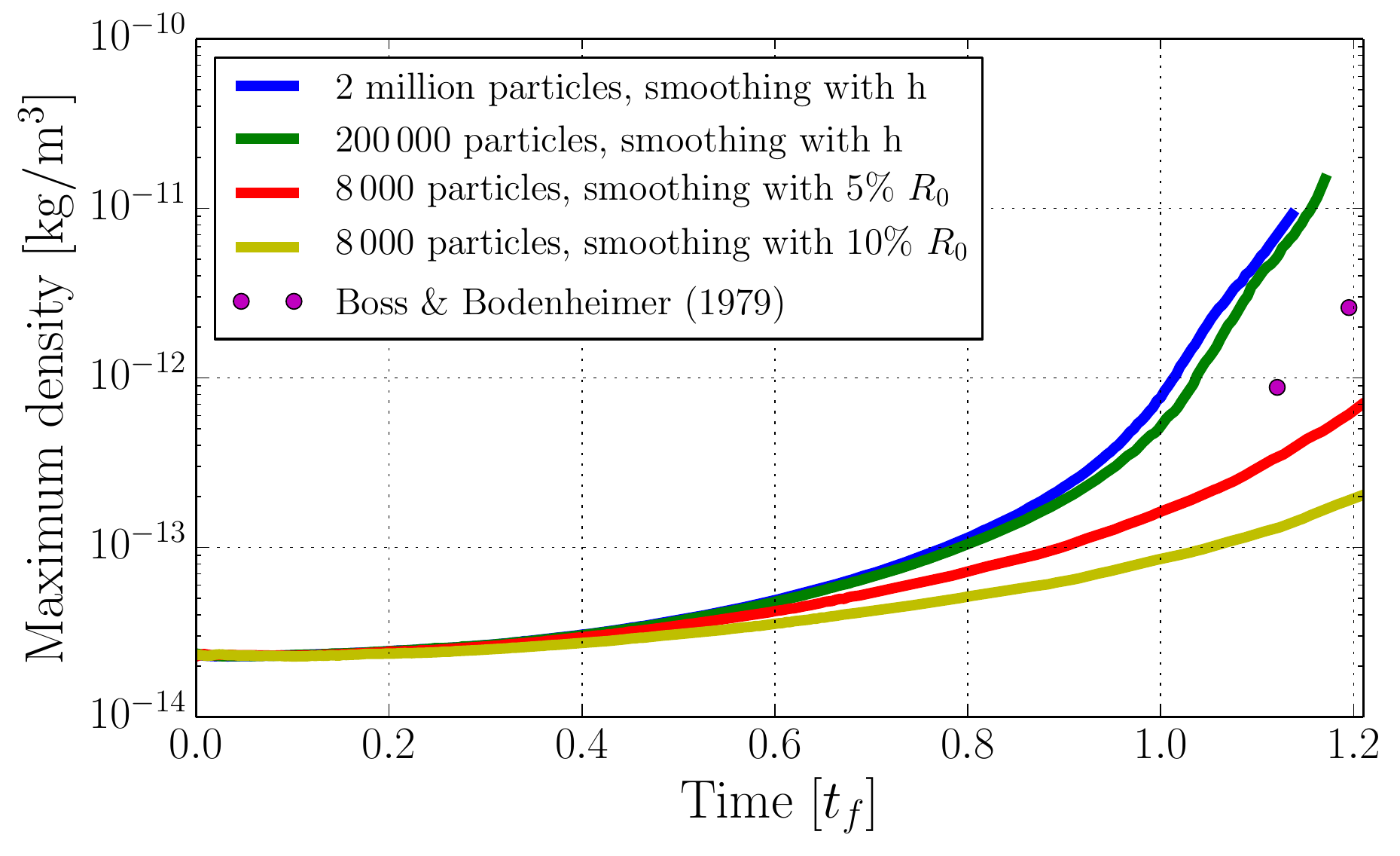}
\caption{Gravitational collapse of a molecular cloud. Evolution of the maximum density for different setups
and smoothing schemes of the gravitational force. The two dots are values by Boss and Bodenheimer from simulations with
grid codes.\label{fig:gravitational_collapse_density}}
\end{figure}

Figure~\ref{fig:gravitational_collapse_density} shows the evolution of the maximum density during the collapse for
different setups and smoothing of the gravitational force. In order to estimate the maximum density, we mapped the
SPH particles on a cubic grid to compare our results to the grid code results by \cite{1979ApJ...234..289B}. Since our
spatial resolution is much better than the resolution from the coarse simulations 30 years ago, we obtain a higher
maximum density. Accordingly, we find a slower evolution of the maximum density in our (very) low resolution simulations
including $\num{8000}$ particles. Another major influence on the density evolution is found in the smoothing of the
gravitational forces. In order to avoid numerical instabilities due to close encounters of two particles,
eq.~(\ref{eq:gravitational_force}) has to be modified and the distance between two bodies is smoothed.
We applied three different smoothing schemes for best comparison to the former publications. The
first smoothing algorithm sets the smoothing length as the lowest possible distance between two particles. This
algorithm was used in the simulation with two million particles. Another ansatz is to add an additional tiny
distance to $r_{pp}$ in (\ref{eq:gravitational_force}) regardless of the separation between two gravitational objects.
The values for this tiny distance employed by \cite{bate:1997} are $\num{5}$\% and $\num{10}$\% of the initial radius of
the molecular cloud. Using these different smoothing methods, we receive the density evolutions presented in
figure~\ref{fig:gravitational_collapse_density}.
\subsection{Impact cratering: Comparison to experiment\label{sect:prater}}
\begin{figure}
\includegraphics[width=0.48\textwidth]{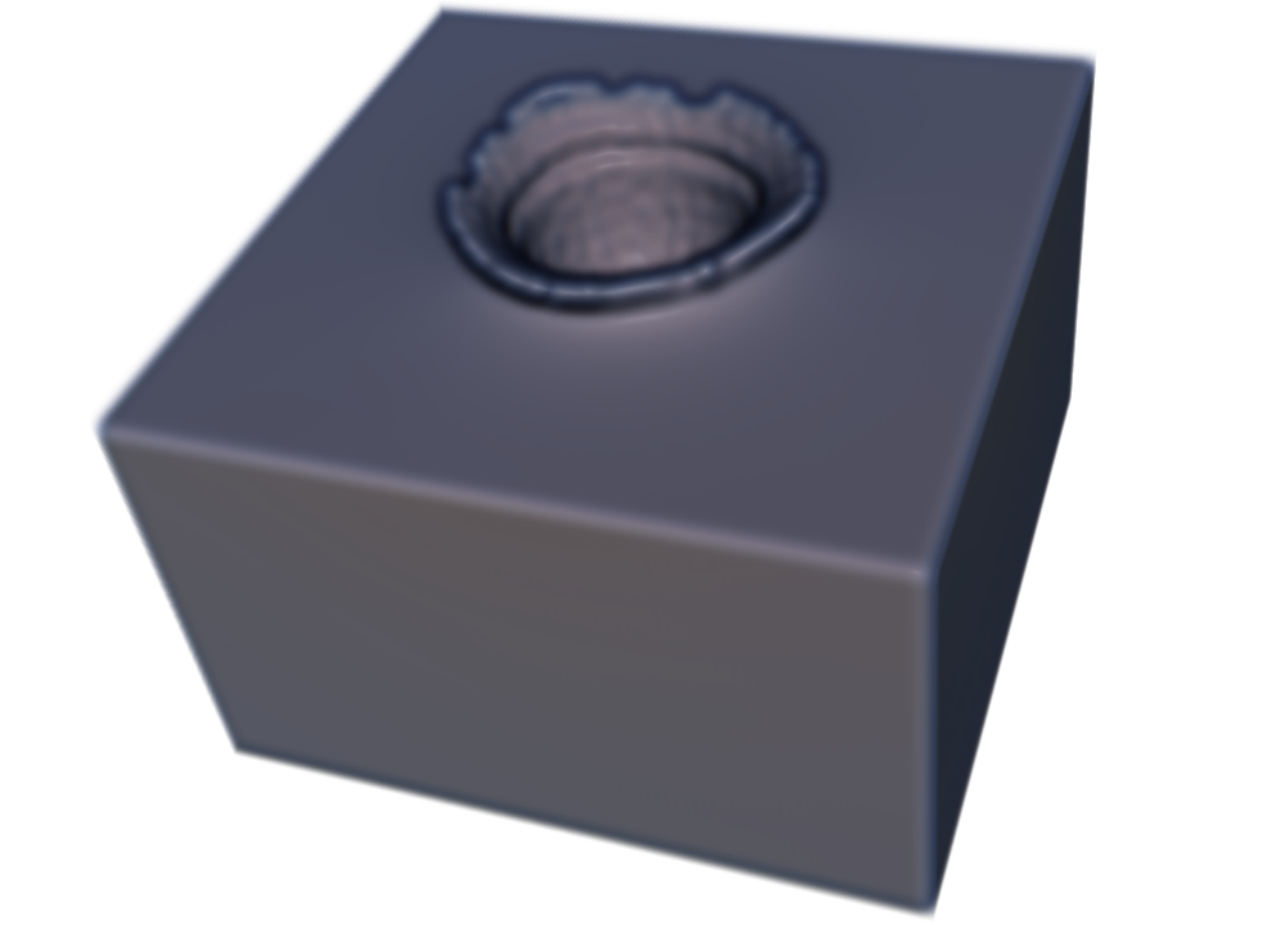}
\caption{Impact into Al6061 simulation. The surface of the cube with the impact crater after 100$~{\mu\mathrm{s}}$ is shown.
The size of the cube is $10 \times 10 \times 10$~$\mathrm{cm}$. The impactor had a radius of ${6.35}~{\mathrm{mm}}$ and the
impact speed was ${7000}~{\mathrm{m/s}}$. 
\label{fig:prater1970}
}
\end{figure}

\begin{figure}
\includegraphics[width=0.48\textwidth]{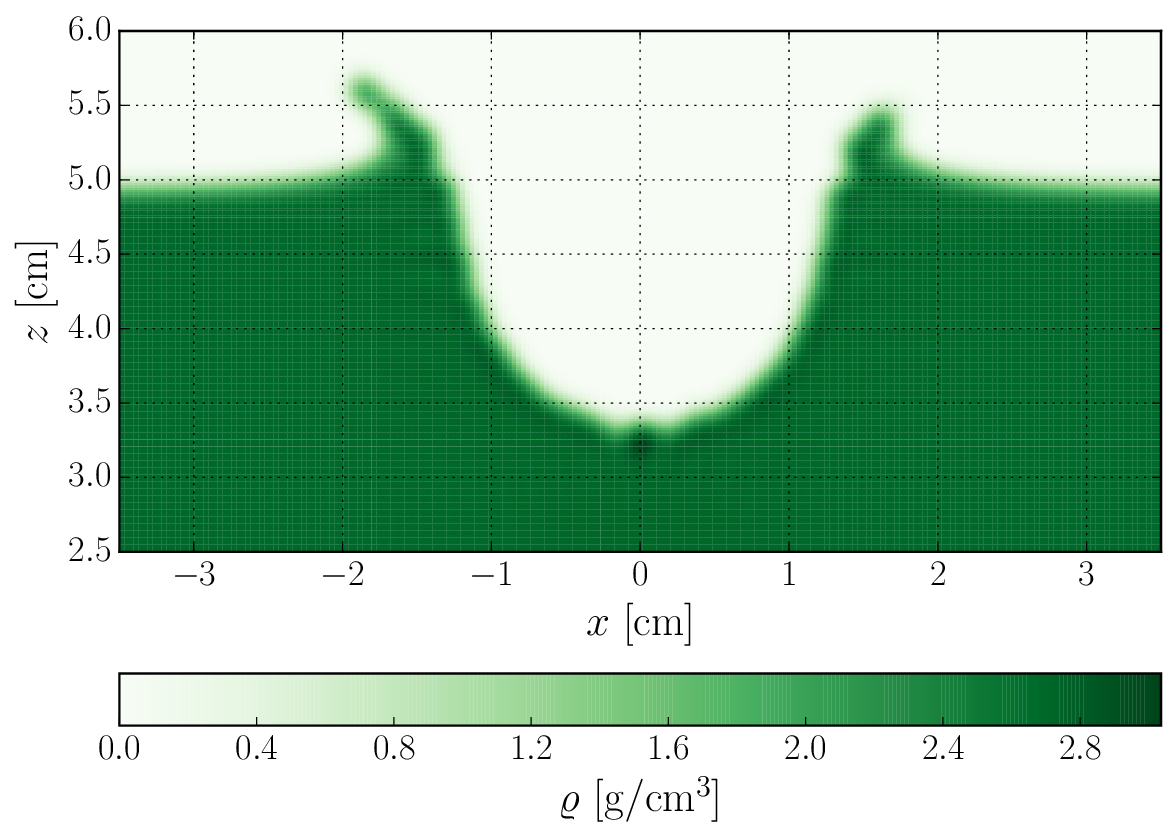}
\caption{Crater dimensions of Al6061 impact simulation. 
The plot shows the color-coded density in a slice through the cube at $y=0$ with the cube edge 
from $-{5}~{\mathrm{cm}}$ to ${5}~{\mathrm{cm}}$ after 100$~{\mu\mathrm{s}}$.
\label{fig:crater_prater1970}
}
\end{figure}

\cite{prater1970hypervelocity} performed several experiments of high-velocity impacts in different aluminium alloys. To validate our code for the use of impact modeling, we  simulated the impact of an aluminium alloy~6061
bullet with radius $\SI{6.35}{\mm}$ in an aluminium alloy~6061 cube with an impact speed of $\SI{7000}{\m \per \s}$. We then
compared our results to \cite{2008M&PS...43.1917P}. 
Since aluminium is not a geological material, this simulation was performed without the use of the Grady-Kipp
damage model, solely with the von Mises plasticity model. 
The three-dimensional simulation has the following initial setup: We
use $\num{500000}$ SPH particles for the cube and $\num{136}$ particles for the impacting bullet, the numerical
parameters are $\alpha=0.5$, $\beta=0$, and the material parameters are $\mu=\SI{26e9}{\Pa}$, $K_0=\SI{75.27e9}{\Pa}$,
$Y_0=\SI{1e8}{\Pa}$, and $\varrho_0=\SI{2.7e3}{\kg \per \cubic\m}$. The parameters for the Tillotson equation of state
$A_T$, $B_T$, $E_0$, $E_{\mathrm{iv}}$, $E_\mathrm{cv}$, $a_T$, $b_T$, $\alpha_T$, $\beta_T$ are
$\SI{75.2e9}{\Pa}$, $\SI{65e9}{\Pa}$, $\SI{5e6}{\J}$, $\SI{3e6}{\J}$, $\SI{13.9e6}{\J}$, $0.5$, $1.63$, $5$ and $5$,
respectively. 
The surface of the cube with the crater after the impact is shown in figure~\ref{fig:prater1970}.
The dimensions of the final crater can be seen in figure~\ref{fig:crater_prater1970}. They are in good agreement with
the numerical results by \cite{2008M&PS...43.1917P} and the experimental results by \cite{prater1970hypervelocity}.
\subsection{Damage model: Comparison to experiment\label{sect:nakamura}}
\begin{figure}
\includegraphics[width=0.48\textwidth]{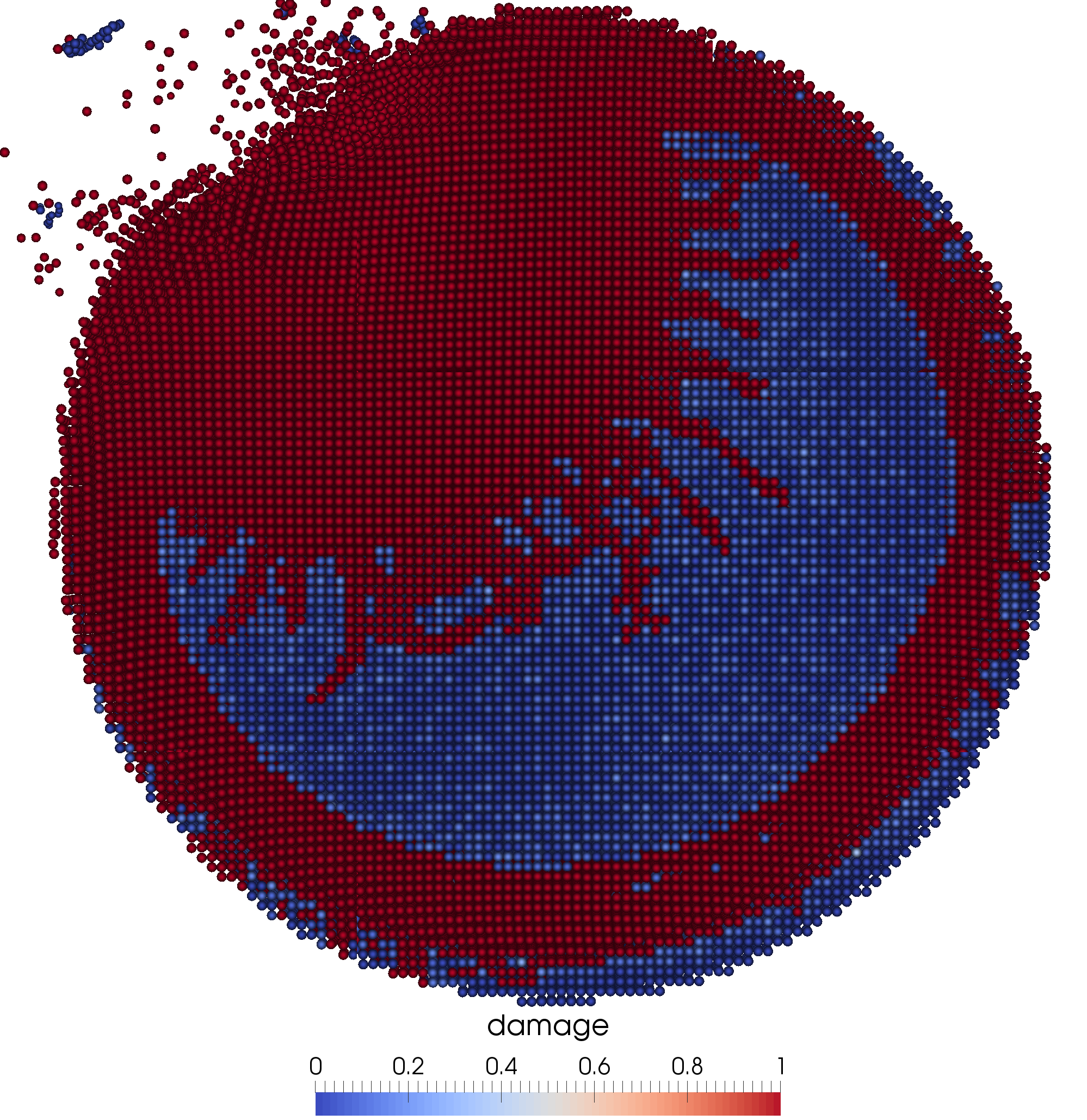}
\caption{\label{fig:nakamura_damage_pattern}
Damage pattern in the basalt target $\SI{50}{\micro\second}$ after the impact of the nylon bullet. The impact
point is at the upper left surface of the spherical target. The diameter of the target is $\SI{6}{\centi\metre}$. The
color code denotes the damage of the particles and the target was split in half to reveal the damage pattern inside. Red particles
are fully damaged. The intact core, indicated by the blue particles, incorporates about $30.4\%$ of the
original target mass.
}
\end{figure}

\begin{figure}
\includegraphics[width=0.48\textwidth]{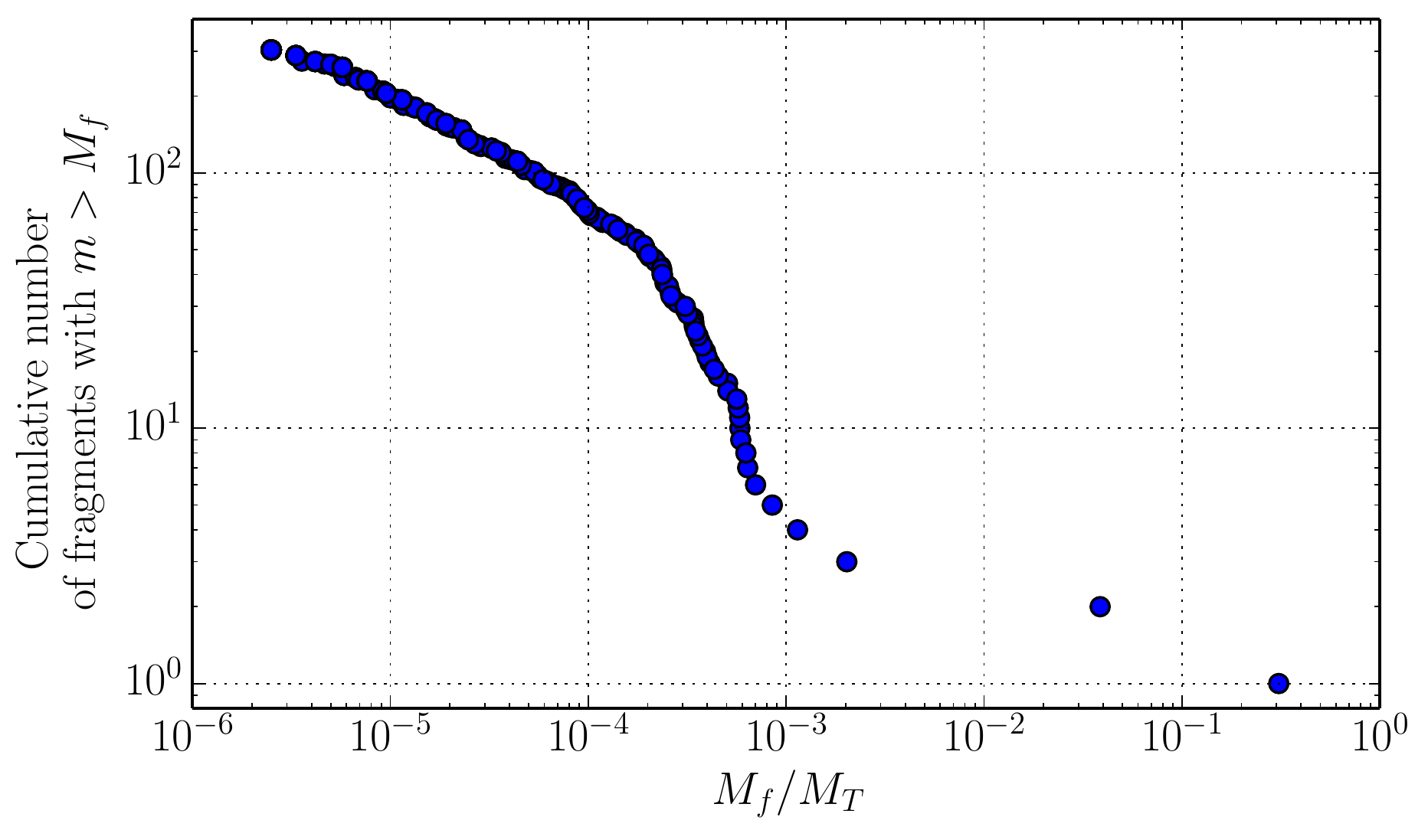}
\caption{\label{fig:nakamura_stats}
Cumulative fragment mass spectrum determined at the end of the impact simulation, calculated by a friend-of-friend
algorithm $\SI{50}{\micro \second}$ after the impact.}
\end{figure}

We simulated a laboratory impact experiment into a basaltic rock performed by \cite{1991Icar...92..132N} to test the implementation of the damage model.
In their experiment, a small $\SI{0.2}{\g}$ nylon bullet impacts with an impact speed of $\SI{3200}{\m\per\s}$ into a
basaltic rocky sphere with a diameter of $\SI{6}{\centi\metre}$. We chose to simulate an off-axis impact with an impact
angle of $\SI{30}{\degree}$ (see fig.~\ref{fig:CeresCollisionGeometry} for the definition of the impact angle), which
corresponds to an impact parameter of 25\% of the diameter. We modeled the basalt target with $\num{523741}$
particles and applied the Tillotson equation of state. Owing to the lack of parameters for nylon, we used lucite for the
projectile material (values were taken from \citep{1994Icar..107...98B}). The basalt target was modeled with the additional
damage model and activation threshold strains were distributed following the Weibull distribution statistically as a
preprocessing step (see sect.~\ref{sect:damage_model} for a detailed description of the algorithm). Following
\cite{1994Icar..107...98B}, we performed several simulations with varying Weibull parameter pairs $k$ and $m$. 
Figure~\ref{fig:nakamura_damage_pattern} shows the damage pattern in the target after $\SI{50}{\micro\second}$ for the
simulation that matched the experimental outcome best. The damage in the region around the impact point is the largest
and all particles close to the impact point were immediately damaged upon impact. The damage pattern is radially symmetric
around the impact point. Another source of damage is found near the surface of the target. There, damage grows due to
spallation stresses that occur when compressive waves get reflected at the free surface. As a result of this nature, the
inner core of the target remains intact while the surface is coated with cracks.
After $\SI{50}{\micro\second}$ simulation time, the damage pattern in the target remains unchanged. We identified
single fragments using a friend-of-friend algorithm in the following way: First, we removed all fully damaged particles.
Then, we clustered all remaining particles that are closer than $1.01$ of the original smoothing length to their
neighboring particles. Fragments consisting of less than three particles were neglected.
Figure~\ref{fig:nakamura_stats} shows the fragment mass spectrum obtained with this friend-of-friend clustering
algorithm.

We find best agreement of the remaining target core size to the experimental data for $k=\SI{5e34}{\per \cubic
\metre}$ and $m=8$, which is close to the values $k=\SI{5e34}{\per \cubic \metre}$ and $m=8.5$ found by
\cite{1994Icar..107...98B}. The intact core incorporates $30.4\%$ of the initial target mass, which has to be compared to
the experimental value of $31\%$.
\section{Collision between Ceres-sized objects and water transport\label{section:application}}
In a forthcoming paper, we study the transfer of water during collisions of Ceres-sized bodies based on preliminary results obtained with our previously presented OpenMP SPH code \citep{maisch13,maidvo14b} and new, higher resolution simulation results gained from the code described here. While \citet{marsas10} show that water contents cannot increase from giant impacts involving bodies with masses between 0.5\,M$_\oplus\/$ and 5\,M$_\oplus\/$, we are interested in possible water delivery by impacts in early planetary systems and hence collision events of smaller bodies, typically involving lower energy and the possibility of water (ice) being transferred between the impactors. Many existing planet formation studies \citep[e.g.,][]{dvoegg12, rayqui04, izides13} assume perfect merging and complete delivery of the asteroids' and planetesimals' water contents to the impact target. As this assumption does not hold, we need to closely investigate the impact process itself  to define conditions under which water is actually transferred rather than  lost during a smaller scale collision. By including a realistic water transfer and loss model in planet formation studies the actual amount of water on the final stages of terrestrial planets may be smaller by a factor of 5--10 \citep{hagray07}.
Most simulations of giant impacts use a strengthless material model \citep[e.g.,][]{canbar13} based on self-gravity dominating the material's tensile strength beyond a certain size of the colliding bodies \citep[400\,m in radius;][]{melrya97}. We study collisions of Ceres-sized objects the size of which is just about on the verge of gravity domination and we expect more realistic results from including solid body dynamics as put forth in sect.~\ref{section:physical_model}.

In this section we present first results from the mentioned higher resolution simulations using the subject CUDA code.

\subsection{Scenario setup}
The collision scenarios include two spherical objects composed of basaltic rock and water ice. Target and projectile have one Ceres mass ($\SI{9.43e20}{\kg}\/$) each. The target has a basaltic core with a 30\,wt-\% water ice mantle and the projectile consists of solid basalt. Owing to the different densities, this configuration results in target and projectile radii of $R_\mathrm{T}=\SI{509}{\km}\/$ and $R_\mathrm{P}=\SI{437}{\km}\/$, respectively. Upon contact, the mutual escape velocity of the two bodies is $v_\mathrm{esc}=\SI{516}{\m\per\s}\/$. Neither of the bodies is rotating at the start of the simulation.

For modeling the materials we use the Tillotson equation of state (see sect.~\ref{section:tillotson}) with material parameters as listed in \citet{benasp99} and our table~\ref{table:tillotson}. The damage model (cf.\ sect.~\ref{sect:damage_model}) is implemented with Weibull parameters, which  were measured for basalt by \citet{2007JGRE..112.2001N}, and for water ice we adopt values mentioned in \citet{lanahr84}; see table~\ref{table:weibull}.

\begin{table*}
	\caption{Tillotson equation of state parameters, shear modulus $\mu\/$, and yield stress $Y_0$ in SI units \protect \citep{benasp99}. We note that $A_\mathrm{T}\/$ and $B_\mathrm{T}\/$ are set equal to the bulk modulus.\label{table:tillotson}}
	\begin{tabular}{lrrrrrrrrrrrr}
        \hline\hline
		\multirow{2}{*}{Material} & \tc{$\varrho_0$} & \tc{$A_\mathrm{T}$} & \tc{$B_\mathrm{T}$} & \tc{$E_0$} & \tc{$E_\mathrm{iv}$}
		& \tc{$E_\mathrm{cv}$} & \tc{\multirow{2}{*}{$a_\mathrm{T}$}}
		& \tc{\multirow{2}{*}{$b_\mathrm{T}$}}
		& \tc{\multirow{2}{*}{$\alpha_\mathrm{T}$}} 
		& \tc{\multirow{2}{*}{$\beta_\mathrm{T}$}}
		& \tc{$\mu$} & \tc{\tstrut $Y_0$} \\
		& [$\si{\kilogram \per \cubic \metre}$] & [GPa] & [GPa] & [$\si{\mega \joule \per \kilogram}$] & [$\si{M\joule \per
        \kilogram}$] & [$\si{\mega \joule \per
        \kilogram}$] & & & & 
		& [$\si{\giga \pascal}$]      & [$\si{\giga \pascal}$] \\
		\hline
		Basalt\tstrut & 2700 & 26.7  & 26.7  & 487 & 4.72 & 18.2  & 0.5 & 1.50 &  5.0 &  5.0
		& 22.7       & \; 3.5 \\
		Ice    & 917 & 9.47 & 9.47 &10 & 0.773 & 3.04 & 0.3 & 0.1 & 10.0 & 5.0 & 2.8        & \; 1 \\
		\hline
	\end{tabular}
\end{table*}

\begin{table}
	\centering
	\caption[]{\label{table:weibull}Weibull distribution parameters for basalt and water ice.}
	\begin{tabular}{llll}
		\hline\hline
		Material & $m$ & $k\/$\,[\si{\per \cubic \meter}] & Reference\tstrut\\
		\hline
		Basalt\tstrut & 16  & $10^{61}$ & \citet{2007JGRE..112.2001N}\\
		Ice & 9.1 & $10^{46}$ &  \citet{lanahr84}\\
		\hline
	\end{tabular}
\end{table}

\begin{figure}
	\centering
	\includegraphics[width=0.6\linewidth]{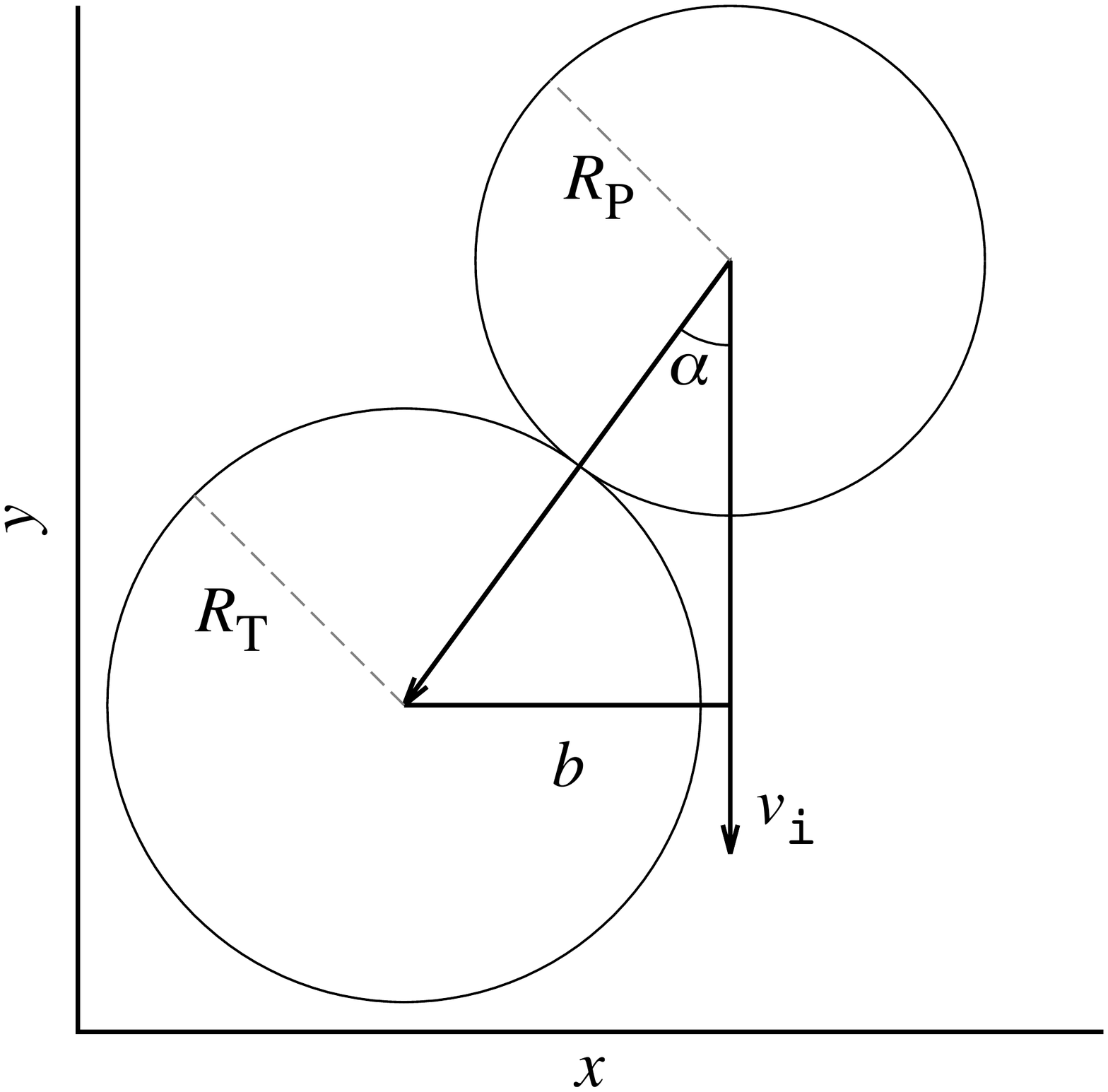}
	\caption{Collision geometry. 
		The collision angle $\alpha\/$ is defined such that $\alpha=0\/$ for a head-on impact. Projectile and target radii are $R_\mathrm{P}\/$ and $R_\mathrm{T}\/$, respectively; $b\/$ is the impact parameter.
		\label{fig:CeresCollisionGeometry}
	}
\end{figure}

Here we present results for a specific hit-and-run collision simulated with a resolution of $\num{200000}$ SPH
particles. The projectile and target start \SI{11369}{\km} apart with an initial impact parameter of $b_0=\SI{819}{\km}\/$ and
a relative velocity of \SI{744}{\m \per \second}. While the individual bodies' SPH particle distributions were relaxed
and in equilibrium before the simulation, we chose this relatively large initial separation of the bodies to allow for a
realistic modeling of tidal effects during their mutual approach. Under the influence of the initial momenta and self-gravity the actual collision happens \SI{234}{\minute} into the simulation. The impact velocity is
$v_\mathrm{i}=1.7\,v_\mathrm{esc}\/$ and the impact angle $\alpha = 48^\circ\/$. The latter is measured such that a
head-on collision corresponds to $\alpha=0\/$; figure~\ref{fig:CeresCollisionGeometry} illustrates the collision
geometry. These impact characteristics are definitely in the hit-and-run domain of the collision outcome map (see e.g.,
\citealt{leiste12}), hence we expect two major survivors that escape each other. In order to be able to clearly identify the surviving bodies and to cover the complete collision process, we simulate for \SI{2000}{\minute} system time.

\subsection{Results}

\begin{figure*}
	\centering
	\fbox{\includegraphics[width=0.48\linewidth]{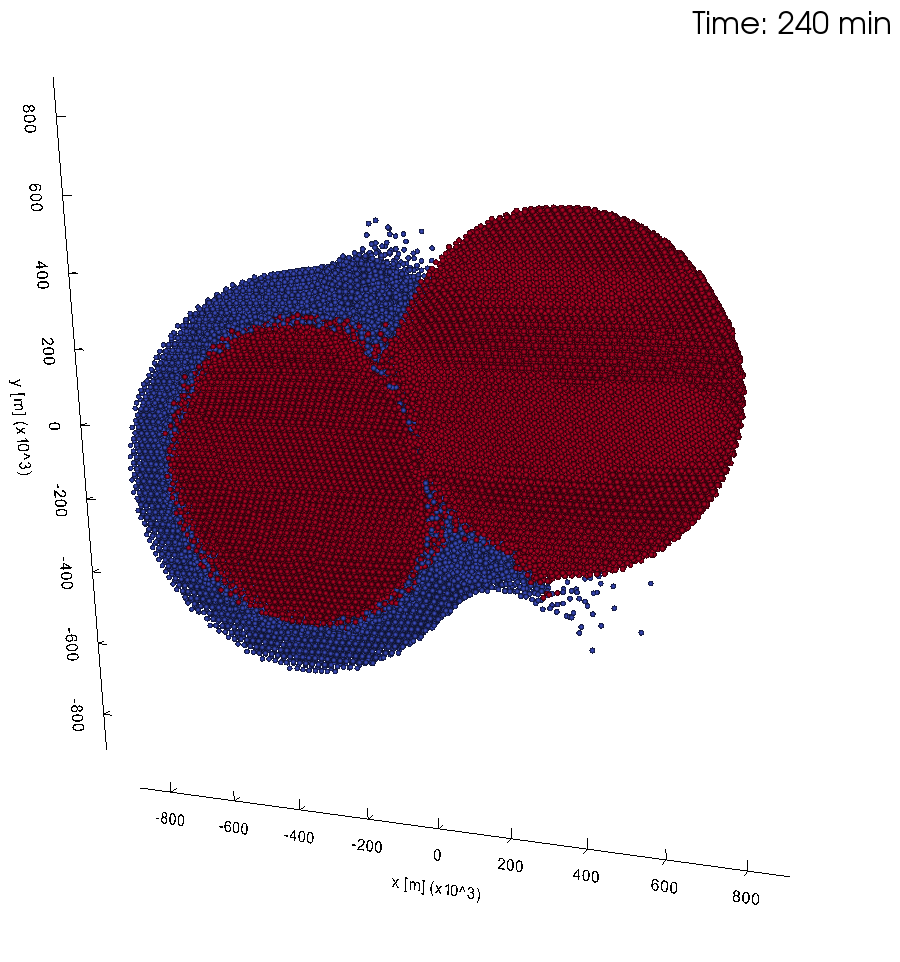}}
	\fbox{\includegraphics[width=0.48\linewidth]{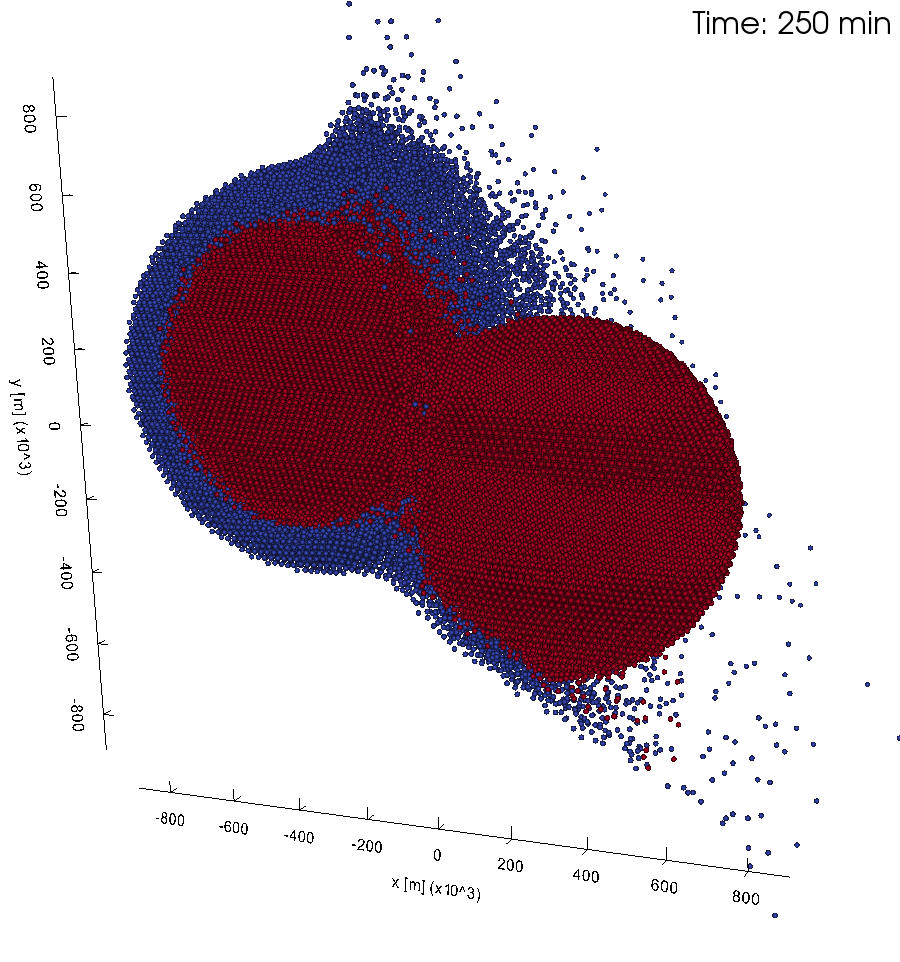}}\\[1ex]
	\fbox{\includegraphics[width=0.48\linewidth]{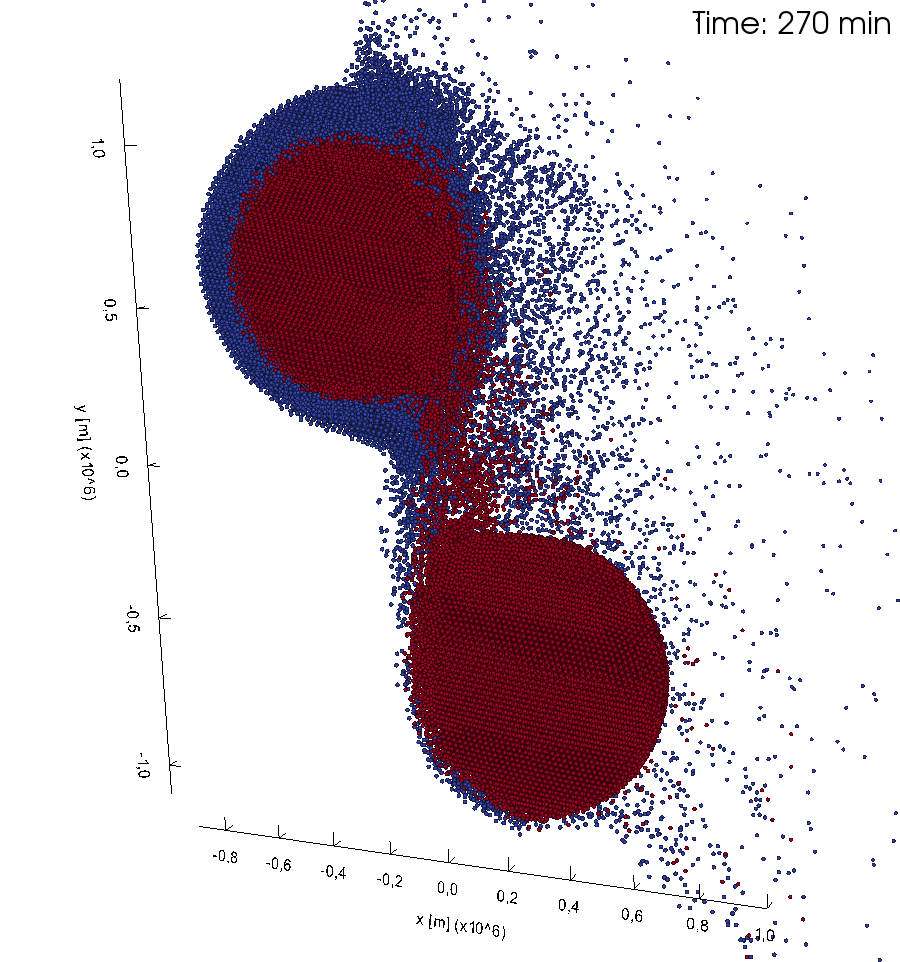}}
	\fbox{\includegraphics[width=0.48\linewidth]{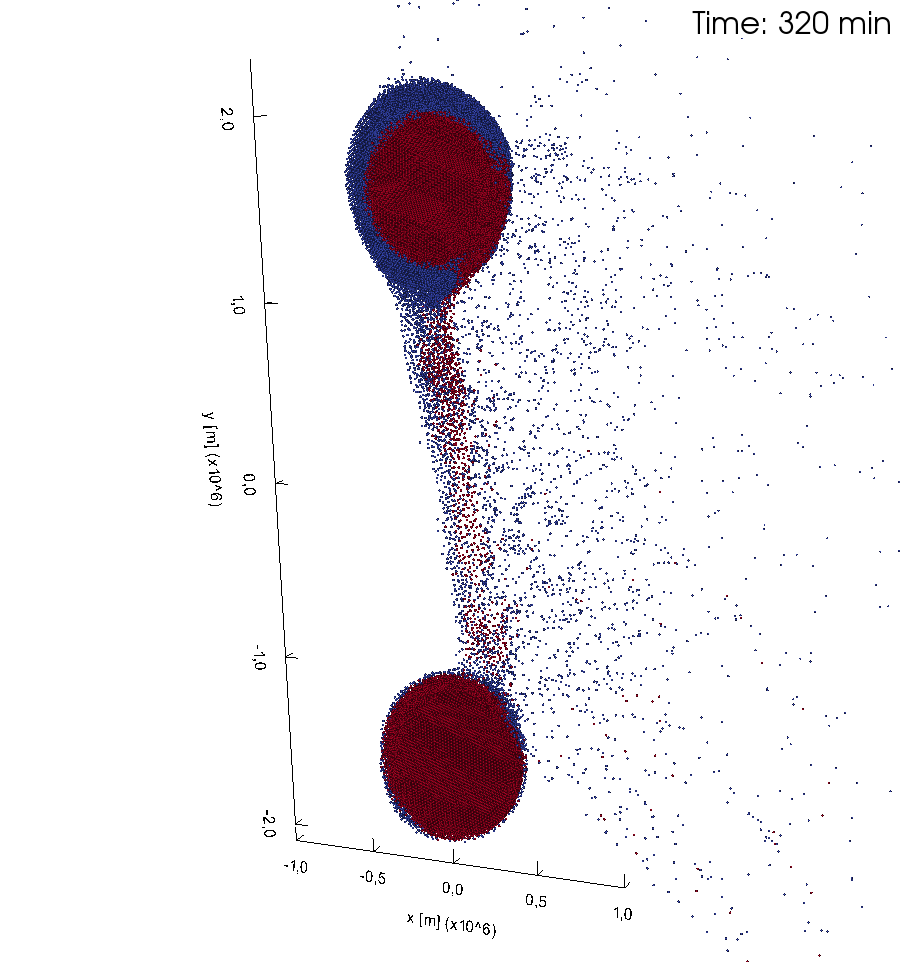}}
	\caption{Collision snapshots. 
		In the right top corner we indicate the timespan into the simulation for each frame. The collision itself occurs at a time stamp of \SI{234}{\minute}. For better visibility of the inside structure we cut open the two Ceres-mass bodies along the $z=0\/$ symmetry plane. Basalt is plotted in red, water ice in blue.
		\label{fig:CeresCollisionSnapshots}
	}
\end{figure*}

\begin{figure}
	\centering
	\fbox{\includegraphics[width=0.9\linewidth]{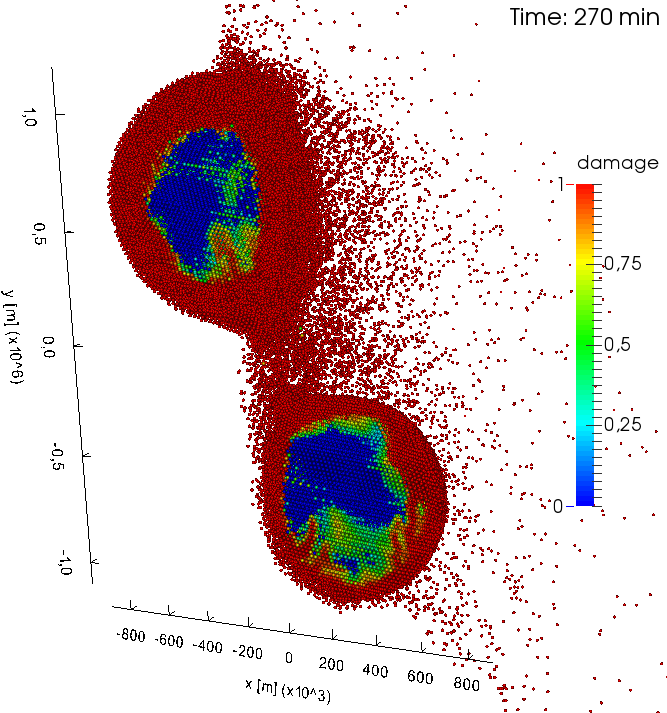}}
	\caption{Damage pattern after the impact (cf.\ the snapshots in fig.~\ref{fig:CeresCollisionSnapshots}). 
                The wet body's icy mantle and outer basalt layers and the surface of the dry body get completely damaged immediately. For better visibility of the inside structure, we again cut the two Ceres-mass bodies along the $z=0\/$ symmetry plane.
		\label{fig:CeresCollisionDamage}
	}
\end{figure}

The simulation results in two surviving bodies of about equal mass that escape each other, as expected for a hit-and-run scenario. Figure~\ref{fig:CeresCollisionSnapshots} shows snapshots from the collision until about \SI{86}{\minute} after, a time when the two bodies are clearly separated again, but still connected by a debris bridge. At the collision site, the icy mantle of the target is stripped and exposes the rocky core. Is it also notable that the initially dry projectile accretes some of the water originally in the target's mantle. The spread-out debris is either reaccreted by one of the two major survivors or it forms fragments.

After the simulation timespan of \SI{2000}{\minute}, the two big objects reach a mutual distance of more than
\SI{60000}{\km}. At this point, we identify the surviving fragments $F_x\/$ using a friends-of-friends algorithm with
the smoothing length as the linking length and find a sharp decrease in mass beyond the two main survivors $F_1\/$ and
$F_2\/$. In total there are 2503 fragments, most of them consisting of one to few SPH particles. In order for $F_k\/$ to
qualify as a ``significant fragment'', we require that it consists of at least 20 SPH particles. Table~\ref{table:frags}
lists the surviving six significant fragments and illustrates the sharp decline in their size beyond the two major
survivors. Given the differing mass of icy and rocky SPH particles by a factor of about three, the number of
particles $n_k\/$ in a fragment is not necessarily proportional to its mass. Beyond the two main bodies, the clumps
consist of just barely over 20 SPH particles so that their properties are not well defined.

Given the $F_1\/$-$F_2\/$ mutual escape velocity at their final separation of \SI{64}{\meter \per \second} the collision outcome is clearly hit-and-run. For the smaller fragments we determined their escape velocities $v_{\mathrm{esc},k}\/$ at their distance from the system's  barycenter $r_k\/$ via
%
\begin{equation}
v_{\mathrm{esc},k} = \sqrt{\frac{2\,G\,M_r}{r_k}},
\end{equation}
\noindent where $M_r\/$ is the aggregated mass of all fragments that are closer to the barycenter than $r_k\/$. For all
$F_{3\ldots 6}\/$ the $v_{\mathrm{esc},k}\/$ values are much smaller than $v_k\/$ (between 5.2 and \SI{6.7}{\meter
\per\second}) so that all of these fragments escape. Further analysis (details not shown) reveals, however, that the two main survivors accrete some of the smaller fragments: $F_1\/$ accretes $F_3\/$, $F_2\/$ accretes $F_5\/$, and $F_4\/$ and $F_6\/$ escape to space.

The small fragments $F_{3\ldots 6}\/$ are originating from debris that is accreted after being ejected by the collision.
Hence, they consist of fully fractured material and are gravitationally bound. The two main survivors' material strength
suffers from the impact but a significant part stays intact; $F_1\/$ is 61\,\% damaged and the originally wet body
$F_2\/$ suffers 70\,\% fracture. Compared to the situation right after the collision, as illustrated in
fig.~\ref{fig:CeresCollisionDamage} (average damage of the two survivors is 64.5\,\%), tidal forces and reaccretion do not significantly add to the degree of overall damage.

\begin{table}
	\centering
	\caption[]{\label{table:frags}Significant fragments  $F_k\/$ after the hit-and-run collision with distances from the center of mass $r_k\/$, number of SPH particles $n_{k}\/$, barycentric velocities $v_k\/$, and wt-\% of water ice $w_k\/$, respectively.}
	\begin{tabular}{llrrrr}
		\hline\hline
		\tstrut & \tc{Mass} & $n_k\/$ & \tc{$r_k$} & \tc{$v_k$} & \tc{$w_k$} \\
		& \tc{[$10^{20}\/$\,\si{\kilogram}]} & & \tc{[\si{\km}]} &  [\si{\meter \per
        \second}] &  [wt-\%] \\
		\hline
		$F_1$ & \num{9.55} & \num{83636} & \num{29801} & \tstrut 278 & 2 \\
		$F_2$ & \num{9.17} & \num{117562} & \num{31113} & 289 & 27 \\
		$F_3$ & \num{0.00252} & \num{33} & \num{20927} & 255 & 29 \\
		$F_4$ & \num{0.00204} & \num{27} & \num{5441}  &  62 & 30 \\
		$F_5$ & \num{0.00177} & \num{34} & \num{18645} & 219 & 66 \\
		$F_6$ & \num{0.00156} & \num{23} & \num{10657} & 120 & 39 \\
		\hline
	\end{tabular}
\end{table}

Concerning the water content, we observe a tendency toward higher values for smaller fragments, but we also notice that
the projectile, which is initially dry, gets a 2\,wt-\% water fraction after colliding with the wet target. There is
also a tendency of the smaller fragments to contain more water than the two main survivors. In total, the surviving
significant fragments have an aggregated mass of \SI{1.87e21}{\kg} and hold \SI{2.73e20}{\kg} ice, which computes to a water fraction of 14.6\,wt-\%.

In summary, we get a big picture result that is consistent with predictions from \cite{leiste12} (hit-and-run outcome), and especially the expected sharply decreasing fragment sizes beyond $F_2\/$. Also, the aggregated water fraction of the significant fragments is in accordance with values obtained via our OpenMP CPU code, which we used to simulate similar impacts with a resolution one order of magnitude lower \citep[\num{20000} SPH particles in ][]{maidvo14b}.

\subsection{Performance\label{sect:applicationperformance}}
We presented a $v_\mathrm{i}=1.7\,v_\mathrm{esc}\/$, $\alpha = 48^\circ\/$ collision of two Ceres-mass bodies. Using a resolution of \num{200000} SPH particles it takes \SI{299}{\hour} wall-time to simulate \SI{2000}{\minute} of the process, consisting of the mutual approach, collision, and subsequent escape of the survivors. Our test was run on a NVIDIA GeForce GTX 770 GPU.

Previously \citep{maidvo14b}, we simulated similar collisions with our OpenMP CPU code using a resolution of only
\num{20000} SPH particles and experienced runtimes of the same order of magnitude for comparable collisions:
\SI{450}{\hour} wall-time for a $v_\mathrm{i}=1.34\,v_\mathrm{esc}\/$, $\alpha = 48^\circ\/$ collision on a dual
Intel Xeon E5410 node and \SI{249}{\hour} wall-time for a
$v_\mathrm{i}=2.13\,v_\mathrm{esc}\/$, $\alpha = 50^\circ\/$ collision on a dual Intel Xeon E5440 node, respectively.

Given the rather similar computing power of the two mentioned CPUs and also given that both the
$v_\mathrm{i}=1.34\,v_\mathrm{esc}\/$ and $v_\mathrm{i}=2.13\,v_\mathrm{esc}\/$ scenarios result in a hit-and-run
collision outcome, the obvious runtime difference might be surprising. It originates from the slower scenario, which is
closer to the merging domain, showing more complex behavior before the two major survivors actually separate.

Even considering these variations from one particular scenario to another, we think there is a clear case for GPU-based code development as we experience a tenfold increase of resolution with about equal processing time when comparing CPU and GPU codes.

\section{Results\label{section:results}}
We have successfully developed a new SPH code that makes use of modern NVIDIA GPUs and includes various different
numerical models. 
The code uses {\tt libconfig} \citep{libconfig} and requires CUDA\texttrademark $5.0$ or higher (the simulations for
this work were performed with CUDA $7.5$). We run the code successfully on various consumer GPUs from NVIDIA, such as  
the GeForce GTX 570, GeForce GTX 680, GeForce GTX 770, GeForce GTX Titan, GeForce GTX TITAN X, and on NVIDIA High Performance Computing Devices such as the
Tesla C2050 / C2070, Tesla K40, and Tesla K80. 
\subsection{Experience with CUDA simulations}
One disadvantage of CUDA is that it is not quite obvious which number of threads should be used to call a certain
kernel. Often, it is necessary to adapt this number to the physical problem or number of particles. Moreover, these
values differ between different GPUs. Hence, sometimes it is more efficient to try a different number of threads for
several kernels before starting a long-time simulation. In our code, the number of threads per kernel may be specified
for the most important kernels during compile time.
\subsection{Runtime comparison: Speed increase\label{sect:results_speedup}}
\begin{figure}
\includegraphics[width=0.48\textwidth]{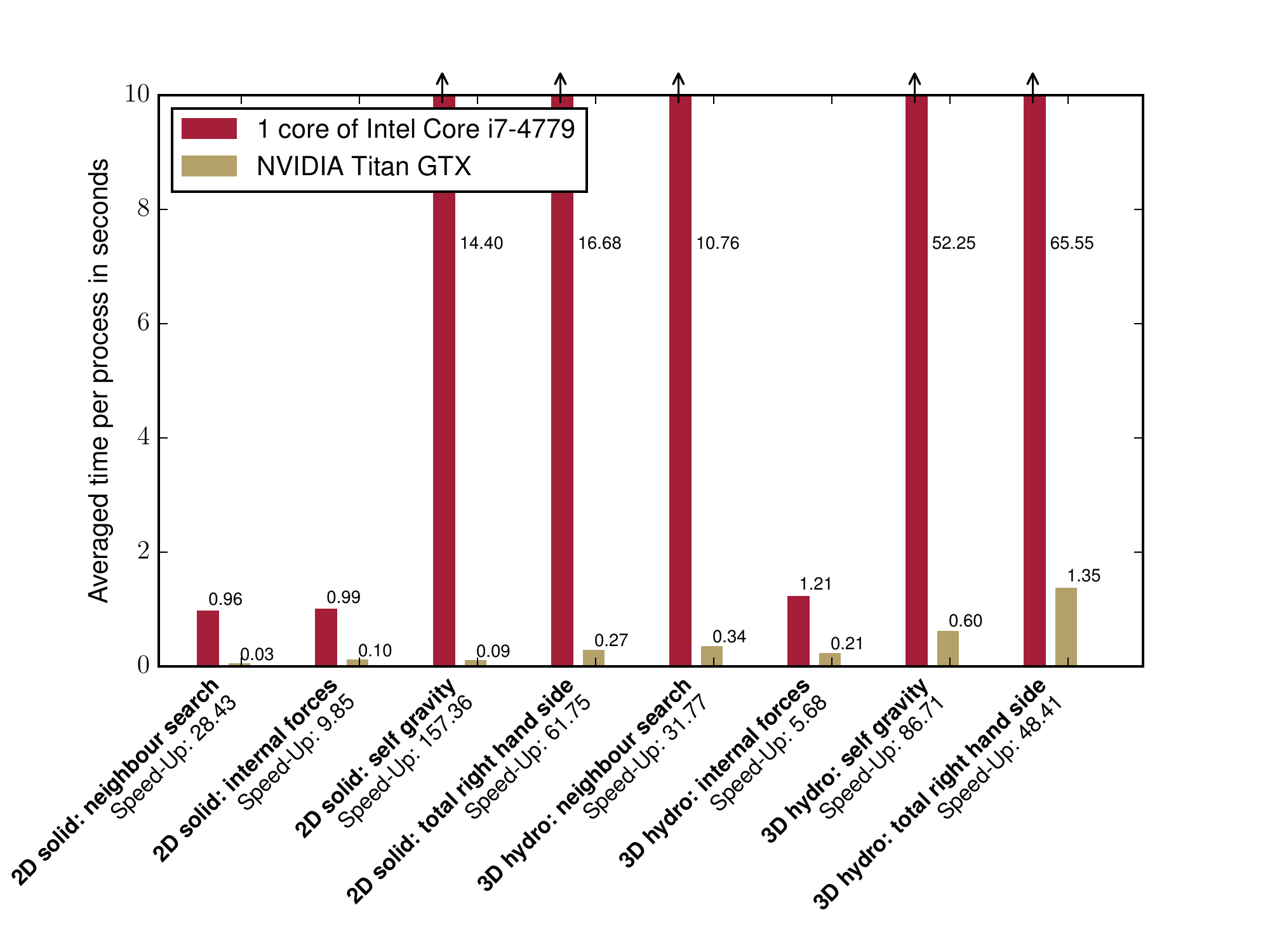}
\caption{Runtime: Nvidia GTX Titan versus single core Intel i7-4779.
\label{fig:speedup}
Comparison between the runtime of typical substeps of the SPH and N-body algorithms for the new CUDA code
and the OpenMP code running on one CPU core. See sect.~\protect \ref{sect:results_speedup} for the description of the substeps.
}
\end{figure}

\begin{figure}
\includegraphics[width=0.48\textwidth]{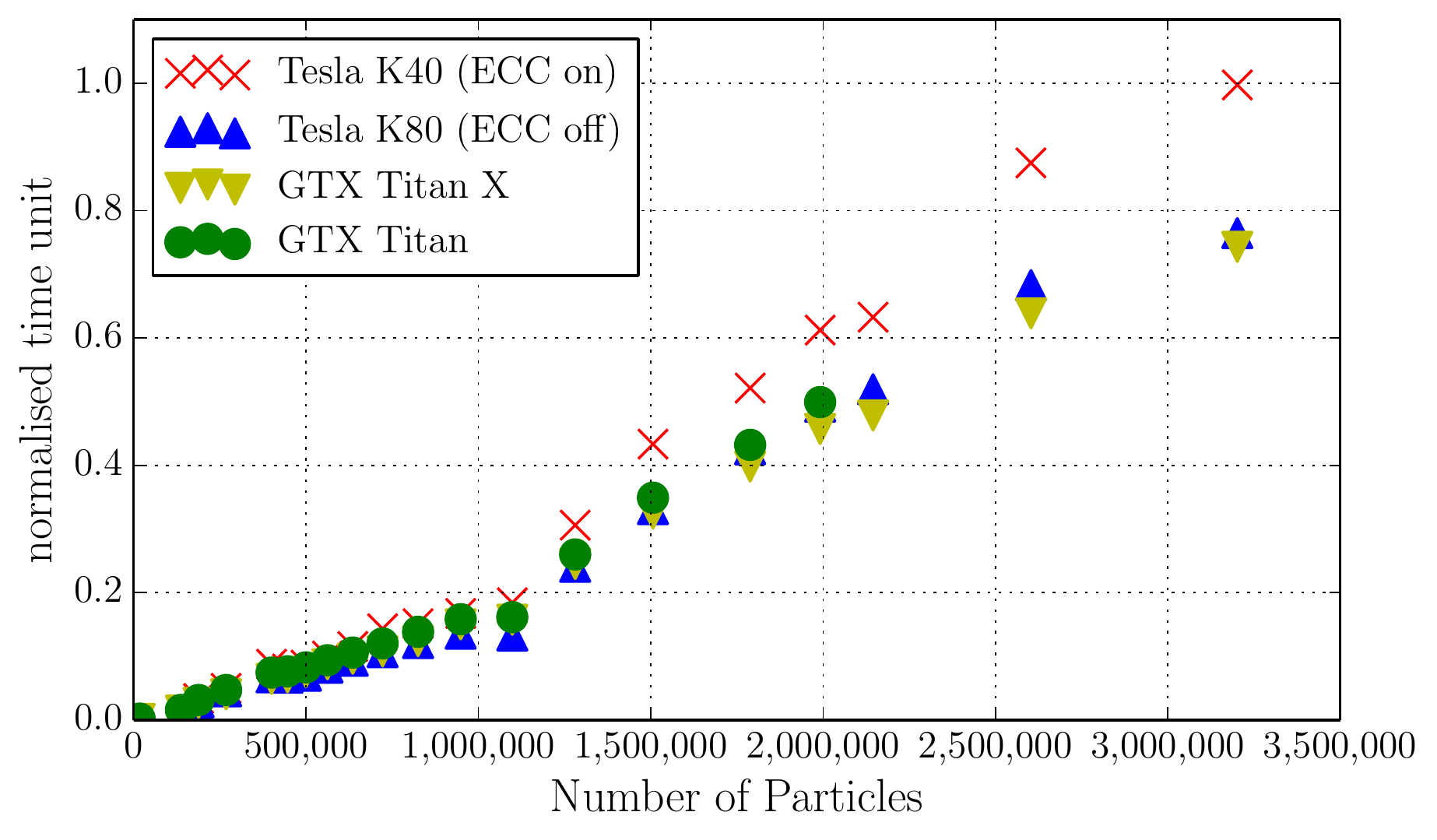}
\caption{Scaling of the code on different GPUs --- Boss-Bodenheimer test case.\label{fig:scaling}}
\end{figure}

Figure~\ref{fig:speedup} shows the main motivation to write a CUDA port of our
SPH code: the performance gain of the implemented algorithms of the GPU compared to a single core
CPU.
We compared the averaged time per substep for different test cases and particle numbers. Of course, these numbers change with different particle numbers and applications. However, we achieve a tremendous increase in speed for the two most
time consuming processes in a SPH simulation including self-gravity: the neighbor search that is used to determine the
interaction partners for all particles and the calculation of the gravitational forces with the Barnes-Hut algorithm.
These two processes add up to more than 80\% of the total computation time. The different subprocesses that are shown
in the plot include the following steps: First, the neighbor search  determines  the interaction partners for all particles.
Second, the internal forces calculates all SPH sums and accelerations from the equation of momentum conservation.
Third, self-gravity includes all computations that are necessary to determine the gravitational forces between the particles. Fourth, the total
right-hand side shows all computations that are necessary to determine the acceleration of all particles.
\begin{figure}
\includegraphics[width=0.48\textwidth]{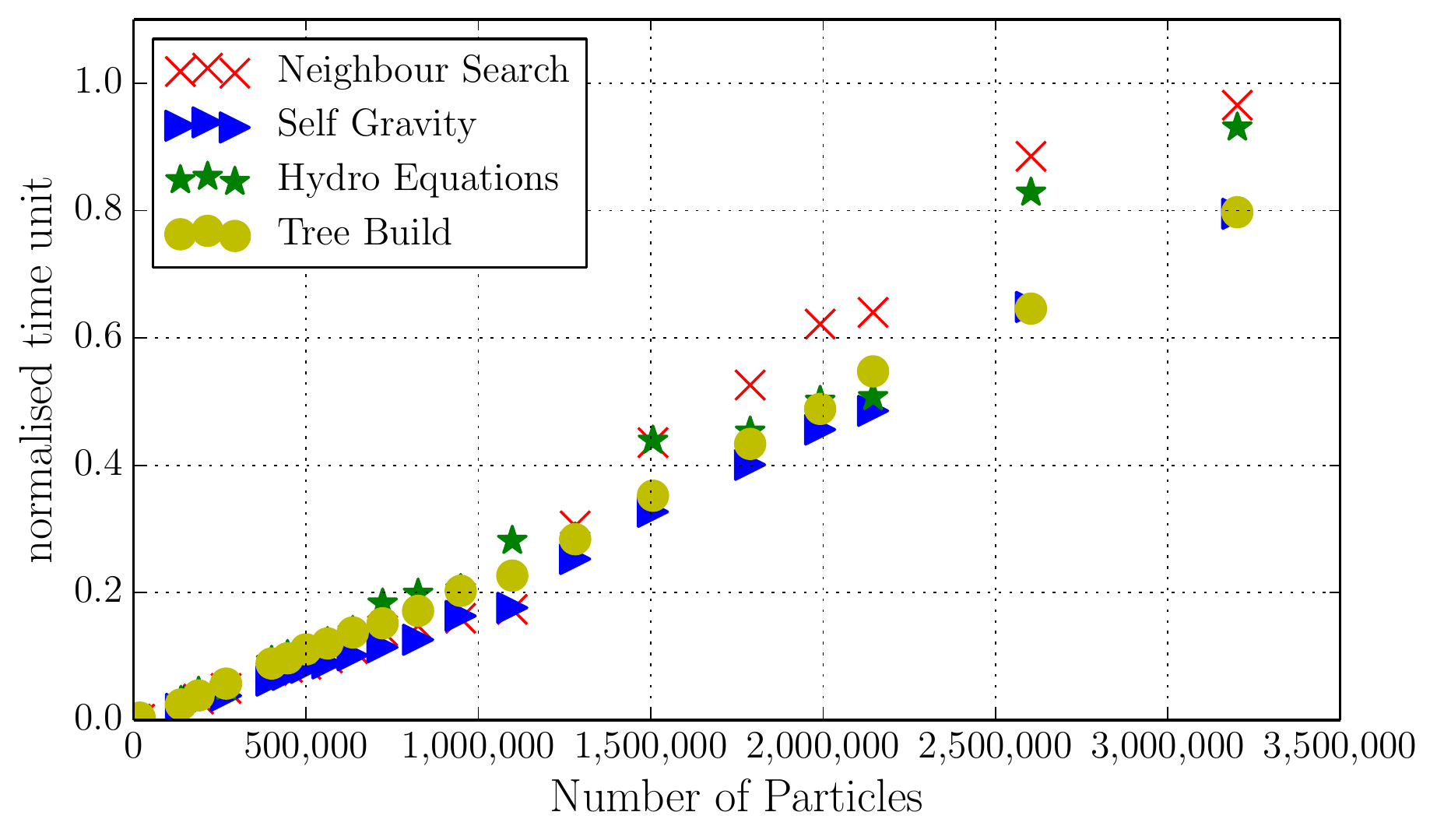}
\caption{Scaling of the different SPH substeps on the NVIDIA Tesla K40 --- Boss-Bodenheimer test case.
\label{fig:k40scaling}
}
\end{figure}
Figure~\ref{fig:scaling} shows the scaling of the code on different GPUs for different number of SPH particles. The time
has been normalized to the time used on the NVIDIA Tesla K40 with the device error-correcting code (ECC) enabled.
Interestingly, the new consumer devices (Maxwell generation) are as fast as the Kepler generation HPC hardware. However,
the consumer GPUs do not provide an ECC mode. The more sophisticated HPC hardware enables the use of ECC for the cost of
20-30\% performance. We have not found any differences in the simulation results that would demand the adoption of an
automatic error correction code. Since we perform all calculations on the GPU and transfer data to the host only for
input/output procedures, the other features of the HPC hardware such as bilateral transfer between host and device are
not extensively relevant.

\begin{figure}
\includegraphics[width=0.48\textwidth]{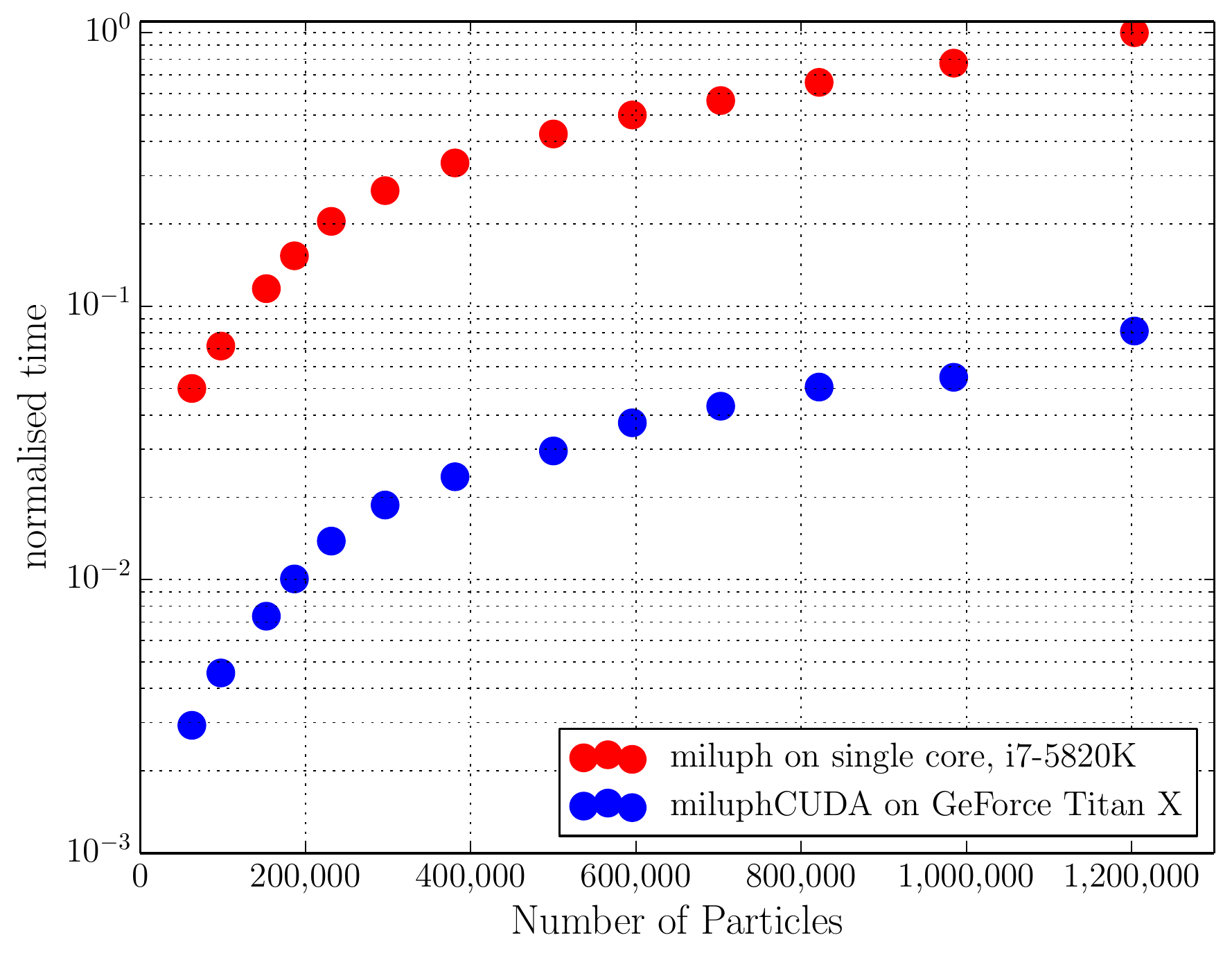}
\caption{Scaling of the code compared to the serial CPU version; impact experiment test case\label{fig:cpuvsgpu}.}
\end{figure}

We additionally analyzed the individual scaling for each subprocess of the numerical scheme.
Fig.~\ref{fig:k40scaling} shows the scaling on the NVIDIA Tesla K40 for different number of SPH particles for the
simulation of the collapse of the molecular cloud. The scheme for the calculation of the self-gravity of the particle
distribution and the scheme for the build of the Barnes-Hut tree show the best scaling. 
The increase in speed of the new CUDA version in comparison to our OpenMP implementation is shown in fig.~\ref{fig:cpuvsgpu}. We
ran the Prater impact experiment (see sect.~\ref{sect:prater}) with both codes for different numbers of SPH
particles. The two codes show similar scaling behavior for the number of particles with a constant increase in speed of the GPU
version. We intentionally used only one core in the OpenMP version runs for better comparison.
In our first astrophysical application of the code, we were able to increase the resolution by
a factor of ten for about equal processing time (see sect.~\ref{sect:applicationperformance}).

\subsection{Numerical limitations}
Modern GPUs currently offer  fast memory in the range up to 12 Gigabytes (e.g.,\ Tesla K40 and K80, GeForce Titan X) and
up to 2880 streaming processors. The size of the memory is the limiting factor for our code. The memory usage strongly
depends on the physical model that is applied. Depending on the specific model and dimension, the code holds
different information for each
particle in the memory. For example, the simplest hydro model only demands information about the location of the
particles, their velocities and their masses, that is eight variables in double precision in 3D and only six variables
in 2D per particle. For the simulation of solid bodies, the number of variables per particle in the memory increases,
since the density, internal energy and deviatoric stress tensor have to be kept available. 
We additionally store the tree and information about the interactions in the memory.
The most
expensive model in regard to memory consumption is the damage model, since we need to save the activation thresholds for
each individual particle in the memory. Hence, the number of required values per particle grows extensively. The mean
number of activation flaws per particle is given by $\log N_p$, where $N_p$ denotes the total number of SPH particles. 
Additionally, the memory demand is related to the choice of the integrator; the Runge-Kutta integrator requires more
substeps as the predictor-corrector scheme to store the additional derivatives at the time of the substeps in memory. 
We  allocated about four gigabytes of memory on the Geforce Titan X for the most demanding simulation in regard to
memory consumption, the Nakamura experiment in
sect.~\ref{sect:nakamura} with $0.52$~million particles. As can be seen in fig.~\ref{fig:speedup}, the most expensive
interaction in terms of computation time is self-gravity. The wall clock time of all simulations including self-gravity is
dominated by the tree algorithm. For some applications it might be helpful to decouple the hydro timescale from the
gravitational timescale, since the code resolves shock waves in the solid bodies while the locations of the particles
hardly change. This allows for reusage of already computed pseudo-particles for the calculation of the accelerations due to self-gravity, as long as the tree properties remain unchanged. We added a command line switch for self-gravity simulations that
enables this feature.
However, we have not used the decoupling for our test simulations presented here.
\subsection{Limitations in terms of physical models}
As was pointed out by \cite{2009P&SS...57..127H}, the von Mises yield criterion does not perfectly describe
the behavior of geologic materials. The fully damaged material after an impact or a collision should not be treated as
a pure fluid, since it behaves more like a granular medium. Hence, the post-impact material flow cannot be modeled
realistically with the von Mises criterion. This has already been investigated by \cite{2015P&SS..107....3J}.
To overcome this limitation, we will implement a proper material model for granular media.
Another limitation is the lack of a porosity model. Numerous astrophysical objects such as comets and asteroids show
high porosities. We would like to apply our newly developed code to simulate collisions between porous objects and
plan to include two porosity models to the code: a strain based model \citep{2006Icar..180..514W, Collins_2011} and a pressure based
model \citep{2008Icar..198..242J}. In the current version of the code, the yield criterion does not depend on the
temperature. The change of temperature because of plastic deformation is not calculated. However, these features have
been already implemented in the CPU version of the code and will be migrated to the CUDA version soon. This will add more
material models to the code, such as the Johnson-Cook plasticity model, where the yield strength depends on temperature
and plastic strain and strain rate.
\section{Conclusions and future work\label{section:conclusion}}
We presented a new SPH code for NVIDIA GPUs using the CUDA framework. The code can be applied to various different
applications in astrophysics and geophysics. It may be used to model hydrodynamics and solid bodies, including ductile
and brittle materials, such as aluminium or basalt. Additionally, the code implements a Barnes-Hut tree that allows for
treatment of self-gravity. The code is freely available to the astrophysical and geophysical community upon request by
email to the authors.

One main advantage of the CUDA implementation is the enormous increase in speed compared to our existing OpenMP SPH code. The
performance of the CUDA code allows for high spatial resolution without the need for cluster hardware. This may enable
researchers without access to HPC hardware to perform sophisticated SPH simulations on workstations. Although the
code runs faster and more efficient on NVIDIA's HPC GPUs such as the Kepler GPUs, a large increase in speed compared to the CPU
version already exists with the use of cheaper consumer graphics cards.
We plan to include more physics in the CUDA version of our SPH code. Currently, we add a porosity model
and a more refined plasticity model to the code.
One major future step will be to add the support for more than one GPU.
\section*{Acknowledgements}
We thank the anonymous referee for the thorough review, positive comments, constructive remarks, and
suggestions on the first
manuscript. Christoph Schäfer wants to thank Daniel Thun for various helpful comments about the mysteries of CUDA.
Most plots in this publication have been made by the use of the matplotlib \citep{Hunter:2007}. Thomas Maindl appreciates support by the FWF Austrian Science Fund project S~11603-N16.
\newline
Xeon is a trademark from Intel Corporation.
GeForce, Tesla, CUDA are trademarks from NVIDIA Corporation.
\begin{longtab}
\begin{longtable}{ll}
\caption{Glossary of symbols.}\\
\hline\hline
\endfirsthead
\caption{Glossary of symbols continued.}\\
\hline\hline
\endhead
\hline
\endfoot
        $a_T,b_T,A_T,B_T,$ & \\ $\alpha_T,\beta_T$ & Parameters for Tillotson EoS \\
        $\alpha,\beta$ & Parameters for artificial viscosity \\
        $c_\mathrm{g}$ & Crack growth speed \\
        $c_\mathrm{s}$ & Sound speed \\
        $C^{\alpha \beta}$ & Correction tensor \\
        $d$ & Dimension \\
        $d_t$ & Depth of tree node \\
        $D$  & Dimension \\
        $E_0$ & Parameter for Tillotson EoS \\
        $E_{\mathrm{cv}}$ & Energy of complete vaporization \\
        $E_{\mathrm{iv}}$ & Energy of incipient vaporization \\
        $\epsilon$ & Scalar strain \\
        $\epsilon_s$ & Parameter for artificial stress \\
        $\epsilon_v$ & Parameter for artificial viscosity \\
        $\dot{\varepsilon}^{\alpha \beta}$ & Strain rate tensor \\
        $\zeta^{\alpha \beta}$ & Artificial stress \\
        $\eta$ & $\varrho/\varrho_0$ \\
        $G$ & Gravitational constant \\
        $h$ & Smoothing length \\
        $J_2$ & Second invariant of $S^{\alpha\beta}$ \\
        $k$ & Parameter for Weibull distribution \\
        $K_0$ & Bulk modulus at zero pressure \\ 
        $m$ & Parameter for Weibull distribution \\ 
        $m_a$     & Mass of particle $a$\\
        $\mu$   & Shear modulus \\
        $M_\odot$ & Solar mass \\
        $n(\varepsilon)$ & Weibull distribution \\
        $n$     & Parameter for artificial stress\\
        $n_\mathrm{act}$ & Number of activated flaws \\
        $n_M$   & Parameter for Murnaghan EoS \\
        $n_\mathrm{tot}$ & Number of total flaws of a particle \\
        $N_\mathrm{children}$ & Number of children of a tree node, $2^D$ \\
        $N_\mathrm{flaws}$ & Number of total flaws \\
        $N_\mathrm{threads}$ & Number of threads on GPU \\
        $N_p$ & Number of SPH particles \\
        $p$     &  Pressure\\ 
        $\Pi_{ab}$ & Artificial pressure for particle $a$ and $b$ \\
        $r_{ab}$ & Distance between particle $a$ and $b$ \\
        $R_0$ & Initial radius of cloud sphere \\
        $R^{\alpha \beta}$ & Rotation rate tensor \\
        $\varrho$ & Density \\
        $\varrho_0$ & Initial density \\
        $s$ & Parameter for kernel function \\
        $s_0$ & Edge length of root node \\
        $s_{d_t}$ & Edge length of node of depth $d_t$ \\
        $S^{\alpha \beta}$ & Deviatoric stress tensor \\
        $\sigma^{\alpha \beta}$ & Stress tensor \\ 
        $t$ & Time \\
        $t_\mathrm{f}$ &  Free-fall time \\
        $T$ & Temperature \\
        $\vartheta$ & Parameter for Barnes-Hut tree \\
        $u$     &  Internal Energy \\ 
        $\chi$  & $\eta - 1$ \\
        $v_\mathrm{esc}$ & Escape velocity \\
        $W(r_{ab};h)$ & Kernel function \\
        $\vect{x_a}$ & Coordinates of particle $a$ \\
        $x_\mathrm{sph}$ & Factor for XSPH algorithm \\
        $\vect{v_a}$ & Velocity of particle $a$ \\
        $Y_0$ & Von Mises yield stress \\
        $\Omega$ & Angular velocity
\end{longtable}
\end{longtab}
\bibliographystyle{aa} 
\bibliography{cp}
\end{document}